\documentclass[aps,prd,twocolumn,nofootinbib,showpacs,superscriptaddress,floatfix]{revtex4-1}  % for review and submission
\usepackage{graphicx}  % needed for figures
\usepackage[utf8x]{inputenc}
\usepackage{dcolumn}   % needed for some tables
\usepackage{bm}        % for math
\usepackage{amssymb}   % for math
\usepackage{amsmath}
\usepackage{comment}
\usepackage{graphicx}
\usepackage{color}
\usepackage{slashed}
\usepackage{url}
\usepackage{hyperref}
\usepackage{lipsum}
\usepackage[normalem]{ulem}
\definecolor{darkturq}{RGB}{0, 206,209}

\newcommand{\Eq}[1]{Eq.(\ref{#1})}
\newcommand{\Eqs}[2]{Eqs.(\ref{#1})-(\ref{#2})}
\newcommand{\Fig}[1]{Fig.\,\ref{#1}}
\newcommand{\Figs}[2]{Figs.\,\ref{#1}-\ref{#2}}

\newcommand{\rt}{\mathbf{r}}

\newcommand{\bt}{\mathbf{b}}
\newcommand{\Pt}{\mathbf{P}}

\newcommand{\as}{\alpha_\mathrm{s}}
\newcommand{\der}{\mathrm{d}}
\newcommand{\gev}{\mathrm{GeV}}

\begin{document}
 
\author{Heikki M\"antysaari}
\email{heikki.mantysaari@jyu.fi}
\affiliation{Department of Physics, University of Jyväskylä, P.O. Box 35, 40014 University of Jyväskylä, Finland}
\affiliation{Helsinki Institute of Physics, P.O. Box 64, 00014 University of Helsinki, Finland}
\author{Niklas Mueller}
\email{nmueller@bnl.gov}
\affiliation{Physics Department, Brookhaven National Laboratory, Bldg. 510A, Upton, NY 11973, USA}
\author{Bj\"orn Schenke}
\email{bschenke@bnl.gov}
\affiliation{Physics Department, Brookhaven National Laboratory, Bldg. 510A, Upton, NY 11973, USA}

\title{Diffractive Dijet Production and Wigner Distributions\\ from the Color Glass Condensate}

\begin{abstract}
Experimental processes that are sensitive to parton Wigner distributions provide a powerful tool to advance our understanding of proton structure. In this work, we compute gluon Wigner and Husimi distributions of protons within the Color Glass Condensate framework, which includes a spatially dependent McLerran-Venugopalan initial configuration and the explicit numerical solution of the JIMWLK equations. We determine the leading anisotropy of the Wigner and Husimi distributions as a function of the angle between impact parameter and transverse momentum. We study experimental signatures of these angular correlations at a proposed Electron Ion Collider by computing coherent diffractive dijet production cross sections in e+p collisions within the same framework. Specifically, we predict the elliptic modulation of the cross section as a function of the relative angle between nucleon recoil and dijet transverse momentum for a wide kinematical range. We further predict its dependence on collision energy, which is dominated by the growth of the proton with decreasing $x$. 
\end{abstract}
\date{\today}
\maketitle
\section{Introduction}
Diffractive processes in deep inelastic scattering of electrons off protons or heavier nuclei can provide important information on the target's structure in coordinate and momentum space \cite{Wusthoff:1999cr}. For example, diffractive vector meson production is sensitive to the spatial profile and fluctuations in the target for the coherent and incoherent cross sections, respectively \cite{Good:1960ba,Miettinen:1978jb,Mantysaari:2016ykx,Mantysaari:2016jaz,Mantysaari:2018zdd}. 

Diffractive dijet production has been argued to provide access to the gluon Wigner distribution of nuclei~\cite{Hatta:2016dxp}. Wigner distributions \cite{Ji:2003ak,Belitsky:2003nz,Lorce:2011kd} encode all quantum information about partons, including information on both generalized parton distributions (GPD) \cite{Diehl:2003ny,Belitsky:2005qn} and transverse momentum dependent parton distributions (TMD) \cite{Collins:1981uw,Mulders:2000sh,Meissner:2007rx,Petreska:2018cbf}. They contain essential information on the partonic spatial and momentum distributions, as well as the distribution of orbital angular momentum inside the nucleon \cite{Aschenauer:2015ata,Smithey:1993zz,Hagiwara:2014iya,Lorce:2011kd,Lorce:2011ni,Lorce:2012jy,Leader:2013jra,Hagiwara:2016kam}. It has further been suggested to use Wigner distributions to quantify entanglement entropy in the proton wave function probed in high energy processes \cite{Kovner:2015hga,Kovner:2018rbf,Calabrese:2004eu,Kharzeev:2017qzs,Hagiwara:2017uaz,Kutak:2011rb,Berges:2017zws,Elze:1994qa,Peschanski:2012cw}. Wigner distributions are quantum distributions and not positive definite, but have a probabilistic 
interpretation in certain semi-classical limits \cite{Moyal:1949sk,Hillery:1983ms,Polkovnikov:2009ys,CKTpaper}.  Other quantum phase space distributions can be constructed, such as Husimi \cite{Husimi} and generalized $P$  distributions, routinely used in quantum optics \cite{Drummond:1980,Drummond:1981}. 

A future Electron Ion Collider (EIC) \cite{Boer:2011fh,Accardi:2012qut,Aschenauer:2017jsk} will allow proton tomography with unprecedented precision, measuring parton position, momentum and spin inside protons and nucleons.
The EIC would open up vast possibilities for understanding the gluon Wigner distribution by measuring diffractive dijet production as a function of the nucleon recoiled momentum and the dijet transverse momentum. Similar research directions may be explored at a possible Large Hadron Electron Collider (LHeC)~\cite{AbelleiraFernandez:2012cc}.

Theoretical descriptions of diffractive cross sections in Deep Inelastic Scattering (DDIS) follow two main approaches: A collinear factorization theorem exists for inclusive diffraction, relying on the existence of a single hard scale to factorize diffractive cross sections into process dependent hard scattering coefficient functions and a set of process independent diffractive parton distribution functions (DPDFs) \cite{Collins:1997sr,Wusthoff:1999cr}. For exclusive processes, like the production of jets in diffractive hadron-hadron collisions, this factorization fails \cite{Collins:1992cv,Collins:2011zzd}. This is because of non-cancellation of soft interactions ('Glauber' or 'Coulomb' gluons) between the incoming hadrons and/or their remnants in the initial and final state. Our main focus in this work is diffractive jet production, previously studied at HERA \cite{Diehl:1994wz,Bartels:1996ne,Bartels:1996tc,Bartels:1999tn,Wolf:2009jm}, where large differences between next-to-leading order calculations
and experimental data from ZEUS \cite{Chekanov:2007rh} and H1 \cite{Adloff:2000qi,Aktas:2006up,Aaron:2010su,Aaron:2011mp,Andreev:2014yra,Andreev:2015cwa} have been attributed to factorization breaking \cite{Kaidalov:2003xf,Klasen:2004qr,Klasen:2008ah}\footnote{We note that the HERA data referred to here is not fully exclusive, allowing for additional (unidentified) particles in the final state (in addition to dijet and proton). Similar processes are considered in ultraperipheral collisions at ATLAS~\cite{ATLAS:2017kwa,Guzey:2018dlm}.}.

Another approach is $k_T$-dependent factorization \cite{Catani:1990eg,Collins:1991ty,Levin:1991ry,Ermolaev:2017qei}  of diffractive cross sections where diffraction occurs via 'Pomeron exchange', which in the perturbative regime can be understood as a two-gluon color singlet exchange. The umbrella term '$k_T$ dependent factorization' encompasses a wide range of approaches, including Balitsky-Fadin-Kuraev-Lipatov (BFKL) evolution \cite{Kuraev:1977fs,Balitsky:1978ic}, the dipole model \cite{Kopeliovich:1981pz,Bertsch:1981py,Mueller:1989st,Nikolaev:1990ja}, the Color Glass Condensate (CGC) \cite{McLerran:1993ni,McLerran:1993ka,McLerran:1994vd,Gelis:2010nm,Albacete:2014fwa},  Balitsky-Kovchegov (BK) evolution \cite{Balitsky:1995ub,Kovchegov:1999yj} and high energy factorization \cite{Catani:1994sq}.

At very small $x$ the Color Glass Condensate effective theory, describing Quantum Chromodynamics (QCD) in the high energy limit, is a suitable framework to compute diffractive processes.
In this framework, slow modes in the fast moving target are highly occupied gluon fields, which can be described classically by the Yang-Mills equations. Fast modes act as sources for these slow modes. Renormalization group equations govern the evolution of the separation between hard and soft modes towards lower momentum fraction. These are the Jalilian-Marian-Iancu-McLerran-Weigert-Leonidov-Kovner JIMWLK equations \cite{JalilianMarian:1996xn,JalilianMarian:1997jx,JalilianMarian:1997gr,Iancu:2001md,Ferreiro:2001qy,Iancu:2001ad,Iancu:2000hn}, which in the limit of a large number of colors simplify to the Balitsky-Kovchegov equations \cite{Balitsky:1995ub,Kovchegov:1999yj}. The JIMWLK equations resum all terms enhanced by large logarithms in $1/x$, resulting in leading logarithmic accuracy. Next-to-leading order corrections were included several years ago \cite{Balitsky:2008zza,Balitsky:2013fea,Kovner:2013ona,Kovner:2014lca}, but, as opposed to the leading logarithmic case \cite{Weigert:2000gi,Rummukainen:2003ns}, at this point no (numerical) solution of the NLO JIMWLK equations has been explored. 

The CGC provides the necessary
ingredients to compute coherent diffraction in the dipole picture. Here, color-singlet Pomeron exchange is understood as the color-diagonal \cite{Good:1960ba} interaction of a dipole with a stochastic ensemble of classical color fields in the eikonal limit. A stochastic average over target color configurations is required, which when performed on the level of the scattering amplitude is equivalent to assuming that the target remains intact ('coherent diffraction').

In the dipole picture, hard diffractive dijet production in small-$x$-deeply inelastic scattering (DIS) has been considered in \cite{Hatta:2016dxp} and in \cite{Altinoluk:2015dpi}. Related work includes~\cite{Boussarie:2018zwg}, where it was suggested to study forward diffractive quarkonia production in $p+p$ collisions to probe the Weizs\"acker-Williams gluon distribution and~\cite{Hagiwara:2017fye}, where access to the gluon Wigner distributions via ultra-peripheral $p+A$ collisions was discussed.  Exclusive double production of pseudoscalar quarkonia and its relation to the gluon Wigner distributions in nucleon-nucleon collisions was explored in \cite{Bhattacharya:2018lgm}. Exclusive diffractive two- and three-jet production in photon-hadron scattering within $k_T$ factorization
and the cancellation of IR, collinear and rapidity singularities for the two-jet cross section was studied in \cite{Boussarie:2016ogo}.

In this manuscript, we present calculations of the gluon Wigner and Husimi distributions in the proton within the CGC framework at leading logarithmic order. We introduce a spatially dependent color charge distribution of the proton, constrained by DIS data from HERA. The energy dependence of the corresponding Wigner distribution is determined by numerically solving the JIMWLK evolution equations. We compute correlations between impact parameter and transverse momentum and extract the elliptic anisotropy coefficient of Wigner and Husimi distributions. 

In order to connect the Wigner distribution to experimental observables, we perform an extensive study of diffractive dijet production cross sections in $e+p$ collision at typical EIC kinematics within the same framework. We focus on dijet kinematics in the so-called correlation limit and predict the dependence of elliptic modulations as a function of the relative angle between nucleon recoil and dijet transverse momentum on collision energy and kinematics, which can be tested at a future EIC.

This manuscript is organized as follows: In Section \ref{sec:WignerIntro} we illustrate how gluon Wigner distributions can be directly computed 
within the CGC in the small $x$-limit. In Section \ref{sec:CSIntro} we discuss the production cross section for dijets in virtual photon-nucleus scattering, for arbitrary photon virtuality and quark mass.
In Section \ref{sec:WIgnerfromCGC} we compute gluon Wigner and Husimi distributions from the CGC and study correlations between
impact parameter and transverse momentum. 

In Section \ref{sec:dijetCS} we present our results for diffractive dijet production cross sections in $e+p$ collisions at typical EIC energies. First, in Section \ref{sec:baselineIPSat} we present
a simple baseline study based on the IP-Sat model and disentangle genuine correlations from kinematic effects. We then compute diffractive cross sections for charm jets from the CGC in Section \ref{sec:fullCGC}
and study the dependence of the elliptic Fourier coefficient on the dijet momentum, photon virtuality and collision energy. 

Our findings are supplemented by multiple appendices: In Appendix \ref{app:detailsWignerCGC} we present details of the
Wigner function derivation from the CGC and a practical approach to numerically compute Wigner and Husimi distributions in this framework. In Appendix \ref{app:defkinematics} we discuss conventions for transverse dijet momentum variables and in Appendix \ref{app:protonsize} we investigate the dependence of azimuthal dijet correlations on the proton size.

\section{Wigner distributions from Coherent Diffractive Dijet Production}
\subsection{The structure of nucleons from Wigner distributions}\label{sec:WignerIntro}
Five dimensional quantum phase space Wigner distributions contain complete information of partons inside a hadron. In this study we focus on gluon distributions in the small $x$ limit and, following the conventions of \cite{Hatta:2016dxp}, we write the gluon Wigner distribution,
\begin{align}\label{eq:definWigner1}
xW(x,\mathbf{k},\mathbf{b}) = \int\frac{\der^2 \Delta}{(2\pi)^2}\, e^{i\boldsymbol{\Delta}\cdot\boldsymbol{b}}\, xG_\text{DP}(x,\mathbf{k},\boldsymbol{\Delta})\,,
\end{align}
as the Fourier transform of the generalized transverse momentum dependent (GTMD) dipole gluon distribution \cite{Meissner:2009ww,Lorce:2013pza,Echevarria:2016mrc},
\begin{align}\label{eq:GTMDgeneral}
x&G_\text{DP}(x,\mathbf{k},\boldsymbol{\Delta}) = 2\int \frac{\der z^- \der^2\mathbf{z}}{(2\pi)^3 \,P^+}e^{-i\mathbf{k}\cdot\mathbf{z}-ixP^+z^-}\nonumber\\
&\times \left \langle P+\frac{\boldsymbol{\Delta}}{2} \left| \text{Tr}\left[ F^{+i}\left(\frac{z}{2}\right)\,\mathcal{U}^{-\dagger} F^{+i}\left(-\frac{z}{2}\right)\,\mathcal{U}^{+}  \right] \right| P-\frac{\boldsymbol{\Delta}}{2}\right\rangle\,.
\end{align}
Here, $F^{+i}\left(\pm z/2\right) = F^{+i}(\pm z^-/2,\pm \mathbf{z}/2)$ is the non-Abelian field strength tensor and $\mathcal{U}^\pm$ are future and past oriented Wilson lines. Bold-faced variables denote 2D transverse coordinates or momenta.
In the CGC-limit \Eq{eq:GTMDgeneral} can be written as\footnote{Details can be found in Appendix \ref{app:detailsWignerCGC}.} \cite{Dominguez:2011wm,Hatta:2016dxp}
\begin{align}\label{eq:CGClimitWIgner}
xG_\text{DP}(x,\mathbf{k},\boldsymbol{\Delta}) = &\frac{2N_c}{\alpha_s}\int \frac{\der^2 \mathbf{x} \der^2 \mathbf{y}}{(2\pi)^4} e^{i\mathbf{k}\cdot \mathbf{r} 
+i \boldsymbol{\Delta} \cdot\mathbf{b}}\nonumber\\&\times \left[\nabla_{x_\perp} \cdot  \nabla_{y_\perp}\right]\, \frac{1}{N_c} \left\langle \text{Tr} U(\mathbf{x})\, U^\dagger (\mathbf{y} ) \right\rangle \,,
\end{align}
where the matrix elements $\langle P+\frac{\boldsymbol{\Delta}}{2}| \dots  | P-\frac{\boldsymbol{\Delta}}{2}\rangle $ are replaced by an average over classical target color configurations.
The properties of the target at small $x$ enter through the energy dependent dipole amplitude, 
\begin{align}\label{eq:dipole}
\mathcal{N}(\mathbf{r},\mathbf{b},x_\mathbb{P})\equiv1-\frac{1}{N_c}\text{tr}\left \langle  U(\mathbf{b}+\frac{\mathbf{r}}{2})\,U^\dagger(\mathbf{b}-\frac{\mathbf{r}}{2}) \right\rangle\,,
\end{align}
where $\mathbf{r}=\mathbf{x} -\mathbf{y} $ is the dipole size and $\mathbf{b}\equiv (\mathbf{x} +\mathbf{y})/2$ the impact parameter. \Eq{eq:dipole} can be directly computed from the fundamental Wilson lines $U(\mathbf{x})$  in the CGC effective theory. Alternatively, one can parameterize \Eq{eq:dipole}, e.g. as in the Golec-Biernat\,-\,W\"usthoff \cite{GolecBiernat:1998js,GolecBiernat:1999qd,GolecBiernat:2001mm,GolecBiernat:2003ym}, or IP-Sat \cite{Kowalski:2003hm} model. 

The dipole amplitude is widely used to compute cross sections of various DIS processes in the small $x$ limit. Within the dipole model, 
proton structure functions \cite{Albacete:2010sy,Rezaeian:2012ji,Lappi:2013zma,Mantysaari:2018nng}, single \cite{Lappi:2013zma,Tribedy:2010ab,Fujii:2013gxa,Ma:2014mri,Ducloue:2015gfa,Ducloue:2016pqr} and double inclusive \cite{Albacete:2010pg,Stasto:2011ru,Lappi:2012nh,Kovchegov:1999ji} particle production in proton-proton collisions, and diffractive processes \cite{kuroda:2005by,Kowalski:2008sa,Lappi:2010dd,Toll:2012mb,Lappi:2014foa,Mantysaari:2016jaz,Mantysaari:2018zdd}, including those in ultra-peripheral nucleus-nucleus and
proton-nucleus collisions \cite{Goncalves:2005yr,Caldwell:2010zza,Caldwell:2009ke,Lappi:2013am,Mantysaari:2017dwh} have been computed. 

Most inclusive processes are not very sensitive to the full momentum and spatial structure of the nucleon, as impact parameter and/or transverse momenta are usually integrated over. In contrast, more exclusive cross-sections can provide access to five-dimensional gluon Wigner distributions and diffractive dijet production is particularly well suited for this task~\cite{Hatta:2016dxp}. To probe impact parameter and transverse momentum dependent gluon dipole distributions, the authors of \cite{Hatta:2016dxp} proposed to study coherent diffractive dijet production off a nuclear target in the \textit{correlation limit} (defined below) -- a process which can be probed at the future Electron-Ion-Collider.
\subsection{Diffractive Dijet Production in the Dipole Picture}\label{sec:CSIntro}
In this section, we outline the basic elements for the calculation of coherent diffractive dijet production in DIS off a nuclear target,
\begin{align}
l(\ell)+N(P)\rightarrow l'(\ell ')+N'(P')+\bar{q}(p_0)+q(p_1)\,,
\end{align}
where $\ell,\ell'$ ($P,P'$) are in- and out-going lepton (target) four-momenta, $q\equiv \ell'-\ell$ and $p_0\equiv (p^+_0,p^-_0,\mathbf{p}_0)$ and $p_1\equiv (p^+_1,p^-_1,\mathbf{p}_1)$ are the four momenta of the the produced dijet. In particular, $p^+_{0}\equiv z q^+$, $p^+_{1}\equiv \bar{z} q^+$ denote the light-cone momenta of the jets, where $q^+$ is the photon longitudinal momentum and $\bar{z}=1-z$. 

The scattering process occurs via the exchange of a virtual photon between
target and lepton, and can be interpreted as photon-nucleon scattering for given lepton kinematics, $\gamma^*(q)+N(P)\rightarrow N'(P')+\bar{q}(p_0)+q(p_1)$.
In the dipole approximation, the scattering then entails the virtual photon fluctuating into a quark-antiquark color dipole, 
which in turn interacts with the color field of the nucleus. In the eikonal approximation the quarks exchange color with the target and receive a 'kick' of transverse momentum.  In coherent diffractive scattering 
no net color is exchanged between projectile and target and the target stays intact. Below, we call outgoing quark and antiquark 'jets', but a fully
phenomenological study requires fragmentation via event generators \cite{Dumitru:2018kuw}. 

Following \cite{Beuf:2011xd,Altinoluk:2015dpi}, we write the S-matrix element for the diffractive process $\gamma^*N\rightarrow \bar{q}q N' $ as 
\begin{align}
\mathcal{S}\equiv \langle &\bar{q}_{f_0,h_0,a,p_0}, q_{f_1,h_1,b,p_1} | \hat{S} | \gamma^*_{\lambda,q}\rangle  \nonumber\\&=-8\pi^2 q_f \sqrt{z \bar{z}} \delta(z+\bar{z}-1)\delta_{f_0,f_1} \delta_{h_0,-h_1}\nonumber\\&\times\int\limits_{\mathbf{x}_0}\int\limits_{\mathbf{x}_1} e^{i\mathbf{p}_0\mathbf{x}_0}e^{i\mathbf{p}_1\mathbf{x}_1}\, \Psi_\lambda(\mathbf{r},Q^2)\,\left[U(\mathbf{x}_0) U^\dagger(\mathbf{x}_1)\right]_{ab} 
\end{align}
where $q_f=Z_f e$, $f_{0,1}$, $h_{0,1},a,b$ are charge, flavor, helicity and color of the outgoing dijets, $U(\mathbf{x})$ Wilson lines in the fundamental representation (discussed below) and $\int_{\mathbf{x}_{0,1}}\equiv\int \der^2 \mathbf{x}_{0,1}$. Here, $\Psi_\lambda(\mathbf{r},Q^2)$
is the photon light-cone wave function with longitudinal or transverse polarization $\lambda=0,\pm 1$ describing the splitting into a dipole of size and orientation $\mathbf{r}=\mathbf{x}_0-\mathbf{x}_1$ \cite{Kovchegov:2012mbw,Beuf:2011xd}.

For the coherent diffractive process, the target remains intact and we average over target color configurations on the amplitude level. 
The production cross section of a dijet pair with momenta $(p^+_{0/1},\mathbf{p}_{0/1})$ can be written as
\begin{align}\label{eq:defcs}
p^+_0p^+_1\frac{\der\sigma}{\der p^+_0 \der p_1^+ \der^2 \mathbf{p}_0\der^2 \mathbf{p}_1} = \frac{\delta(p_0^++p_1^+-q^+)}{(2\pi)^5 2|q^+|} \Big|\langle \mathcal{M}\rangle\Big|^2\,,
\end{align}
where $\mathcal{M}$ is the scattering amplitude, defined as $\mathcal{S}=\mathcal{I}+(2\pi) |q^+|\delta(p_0^++p_1^+-q^+) i \mathcal{M}$, with $\mathcal{S}$ the S-matrix. 
A diffractive process is characterized by color neutral exchange ('Pomeron-exchange') and the resulting rapidity gap
between the outgoing dijet system and the struck target, where no particles are produced. The 
scattering matrix is color-diagonal and allows us to write the dipole-target interaction as \cite{Good:1960ba}
\begin{align}
\left[U(\mathbf{x}_0) U^\dagger(\mathbf{x}_1)\right]_{ab}\rightarrow \frac{1}{N_c}\text{tr}\left[U(\mathbf{x}_0) U^\dagger(\mathbf{x}_1)\right] \,\delta_{ab}\,.
\end{align}
In this study, we express dijet production cross sections in the transverse momentum variables\footnote{An alternative coordinate choice is discussed in Appendix \ref{app:defkinematics}.}
\begin{align}\label{eq:CTRvariables}
\mathbf{\Delta}&\equiv\mathbf{p}_0+\mathbf{p}_1\,,\nonumber\\
\mathbf{P}&\equiv\frac{1}{2}(\mathbf{p}_0-\mathbf{p}_1)\,.
\end{align}
In these coordinates, the dijet production cross section for transversely polarized photons reads \cite{Beuf:2011xd,Altinoluk:2015dpi}
\begin{align}\label{eq:transvCS}
&\frac{p^+_0p^+_1\,\der\sigma_T}{\der p^+_0 \der p^+_1 \der^2 \mathbf{\Delta}\der^2 \mathbf{P}}  = \frac{2N_c \alpha_{EM} Z_f^2}{(2\pi)^6}\,\delta(p_0^++p_1^+-q^+) |q^+| \,z\bar{z}\nonumber\\
\times& \Big\{ \zeta^2 \sum\limits_\lambda
\Big| \int\limits_\mathbf{b}\int\limits_\mathbf{r} e^{i\mathbf{b}\cdot\mathbf{\Delta}+i\mathbf{r}\cdot\mathbf{P}}\, \frac{\boldsymbol{\epsilon}_\lambda(q)\cdot \mathbf{r}}{|\mathbf{r}|}
\bar{Q}\,K_1(\bar{Q}|\mathbf{r}|)\, \mathcal{N}(\mathbf{r},\mathbf{b},x_\mathbb{P})\Big|^2 
\nonumber\\&\quad+m_q^2\Big| \int\limits_\mathbf{b}\int\limits_\mathbf{r}e^{i\mathbf{b}\cdot\mathbf{\Delta}+i\mathbf{r}\cdot\mathbf{P}}\, K_0(\bar{Q}|\mathbf{r}|) \, \mathcal{N}(\mathbf{r},\mathbf{b},x_\mathbb{P})\Big|^2 
\Big\}\,,
\end{align} 
where $\mathcal{N}$ is given by \Eq{eq:dipole} and we abbreviated $\bar{Q}\equiv(z\bar{z}Q^2+m_q^2)^{1/2}$, $\alpha_{EM}=e^2/4\pi$ and $\zeta^2\equiv  z^2+\bar{z}^2$.  Here, $\boldsymbol{\epsilon}_\lambda(q)=1/\sqrt{2}(1,\lambda i)$ are light-cone photon polarization vectors and $K_{0,1}$ are modified Bessel functions of the second kind.
The longitudinal momentum fraction of the 'Pomeron' is defined by
\begin{align}\label{eq:xPomeron}
x_\mathbb{P}&\equiv \frac{M^2+Q^2-t}{W^2+Q^2-m_N^2}\nonumber\\
&=\frac{\frac{1}{z\bar{z}}\left(m^2_q + \frac{1}{4} \mathbf{\Delta}^2 +\mathbf{P}^2 + [\bar{z}-z]  \mathbf{\Delta}\cdot \mathbf{P}\right)+Q^2}{W^2+Q^2-m_N^2}\,,
\end{align} 
where $Q^2\equiv -q^2=-(\ell-\ell')^2$ is the virtuality of the photon, $W^2=(P+q)^2$  the center-of-mass energy squared of the photon-target system, $m_N$ the target mass,
$M^2$  the invariant mass-squared of the dijet system and $t\equiv-(P'-P)^2=-\mathbf{\Delta}^2$ is the Mandelstam variable.% 

Analogously, the cross section for photons with longitudinal polarization reads
\begin{align}\label{eq:longtCS}
&\frac{p^+_0p^+_1, \der\sigma_L}{\der p^+_0\der p^+_1  \der^2 \mathbf{\Delta}\der^2 \mathbf{P}}  = \frac{8 N_c\alpha_{EM} Z_f^2 }{(2\pi)^6}\,{Q}^2 \,|q^+|\delta(p_0^++p_1^+-q^+) \nonumber\\ &\qquad\qquad\times(z\bar{z})^3\,\Big| 
\int\limits_\mathbf{b}\int\limits_\mathbf{r}e^{i\mathbf{b}\cdot\mathbf{\Delta}+i\mathbf{r}\cdot\mathbf{P}}  K_0(\bar{Q}|\mathbf{r}|) \, \mathcal{N}(\mathbf{r},\mathbf{b},x_\mathbb{P})  \Big|^2\,.
\end{align}
Angular correlations can be
parameterized by azimuthal Fourier decomposition,
\begin{align}\label{eq:CSdecompostion}
\frac{\der\sigma_{T/L}}{\der\Omega} \equiv \frac{p^+_0p^+_1\,\der\sigma_{T/L}}{\der p^+_0 \der p^+_1 \der^2 \mathbf{\Delta}\der^2 \mathbf{P}} = v_0(1+2 v_2 &\cos[2\theta(\mathbf{P},\mathbf{\Delta})]\nonumber\\& + \dots )\,.
\end{align}
Here, 
\begin{align}
\theta(\mathbf{P},\mathbf{\Delta})\equiv \theta(\mathbf{P})-\theta(\mathbf{\Delta})
\end{align}
is the relative angle between dijet and target recoil momentum.
The coefficients $v_n$ are experimentally measurable by analysis of the azimuthal distribution of dijets. By Fourier transform,  the angular correlation in the dijet cross section is sensitive to the relative orientation between $\mathbf{r}$ and $\mathbf{b}$. 

Here, we study the dijet cross section in the correlation limit, $|\mathbf{P}|\gg |\mathbf{\Delta}|$, where the individual jets are almost back-to-back.
In this limit, one is most sensitive to dipole contributions from the peripheral region at large $|\mathbf{b}|$. In~\cite{Hagiwara:2017fye} it was further shown that in the correlation limit for $Q^2=0$, a direct relation between the diffractive dijet cross section and the gluon Wigner distribution can be established.
\subsection{The dipole amplitude at small $x_\mathbb{P}$}\label{sec:models}
In this section, we discuss two approaches to compute the dipole amplitude, \Eq{eq:dipole}. The first will be a model parameterization from the impact parameter dependent saturation (IP-Sat) model \cite{Kowalski:2003hm},
which has been successfully used to describe a range of data, from HERA inclusive and diffractive $e+p$
DIS data \cite{Rezaeian:2012ji,Rezaeian:2013tka,Mantysaari:2018nng} to n-particle multiplicity distributions in $p+p$ and $p+A$ collisions at RHIC and LHC \cite{Tribedy:2010ab,Tribedy:2011aa,Albacete:2013ei}. The second one will be a CGC computation, where the dipole interaction with the target is
directly encoded in fundamental Wilson lines. 

\subsubsection{IP-Sat model}\label{sec:IPSatmodel} 
In the IP-Sat model the dipole amplitude is given by
\begin{align}\label{eq:IPSat}
\mathcal{N}(\mathbf{r},\mathbf{b},x_\mathbb{P})\equiv 1-\exp{\left\{ \frac{-\pi^2}{2N_c} \mathbf{r}^2 \, \alpha_s(\mu^2)\, x_\mathbb{P}g(x_\mathbb{P},\mu^2)\, T_p(\mathbf{b}) \right\}} \,.
\end{align}
Its impact parameter dependence arises from the transverse spatial color profile of the proton, assumed to be Gaussian,
\begin{align}\label{eq:IPSATprofile}
T_p(\mathbf{b})=\frac{1}{2\pi B_p}\exp{\left\{-\frac{\mathbf{b}^2}{2B_p}\right\}}\,.
\end{align}
The IP-Sat model is $x_\mathbb{P}$-dependent through the Dokshitzer-Gribov-Lipatov-Altarelli-Parisi (DGLAP) \cite{Gribov:1972ri,Gribov:1972rt,Altarelli:1977zs,Dokshitzer:1977sg} evolved gluon distribution function $ x_\mathbb{P}g(x_\mathbb{P},\mu^2)$ at scale $\mu^2 = \mu_0^2 +4/\mathbf{r}^2$.
The proton width $B_p$, the initial scale $\mu_0$ and the initial conditions for $x_\mathbb{P}g(x_\mathbb{P},\mu^2)$ are obtained from fits to HERA DIS data \cite{Mantysaari:2018nng}.
Note that, independent of $|\mathbf{b}|$, the IP-Sat model always parameterizes the long distance behavior as $\mathcal{N}\rightarrow 1 $ for $|\mathbf{r}|\rightarrow \infty$, 
in contrast to the CGC computation for finite systems \cite{GolecBiernat:2003ym,Schlichting:2014ipa}.  

The IP-Sat model is not useful when studying azimuthal correlations in diffractive dijet production, because it does not depend on $\mathbf{r}\cdot\mathbf{b}$, but only on $\mathbf{r}^2$ and $\mathbf{b}^2$ separately. Consequently, it cannot reproduce
any angular correlation between $\mathbf{P}$ and $\mathbf{\Delta}$. In addition, the IP-Sat model contains only DGLAP evolution and we expect it cannot capture important features of the full JIMWLK CGC computation.
However, by artificially introducing angular correlations, we can consider the IP-Sat parameterization as a simple baseline test,
\begin{align}\label{eq:IPSatwithcorr}
\mathcal{N}_C(&\mathbf{r},\mathbf{b},x_\mathbb{P}) \nonumber\\&\equiv 1-\exp{\left\{ \frac{-\pi^2\mathbf{r}^2}{2N_c}  \, \alpha_s(\mu^2)\, x_\mathbb{P}g(x_\mathbb{P},\mu^2)\, T_p(\mathbf{b}) C_{\theta(\mathbf{b},\mathbf{r})} \right\}} \,,
\end{align}
where we included a $\theta(\mathbf{r},\mathbf{b}) \equiv \theta(\mathbf{r})-\theta(\mathbf{b})$ dependent term
\begin{align}
C_{\theta(\mathbf{b},\mathbf{r})}\equiv 1 -\tilde{c}\left[\frac{1}{2}-\cos^2[\theta(\mathbf{r},\mathbf{b})] \right]\,.
\end{align}
Varying the parameter $\tilde{c}$ allows one to regulate the amount of anisotropy in the gluon Wigner distribution by hand and to study the effects on dijet cross sections. Similar parameterization are used in \cite{Altinoluk:2015dpi}, and a more sophisticated analytic expression, albeit with no $x_\mathbb{P}$ dependence, was discussed for example in~\cite{Iancu:2017fzn}. In the latter study, in accordance with our expectations, $\cos [2\theta(\mathbf{r},\mathbf{b})]$ correlations occur in connection with  
gradients of the target color charge density. Specifically, these correlation vanish in the homogeneous regime at small $\mathbf{b}$ and $\mathbf{r}$, a feature not reproduced by  our \Eq{eq:IPSatwithcorr}.

Instead of performing further modifications of the IP-Sat model, we set out to compute the gluon dipole distribution directly from the Color Glass Condensate effective theory at small $x_\mathbb{P}$, including its energy evolution via the JIMWLK renormalization group equations in the next subsection.

\subsubsection{Color Glass Condensate computation}\label{sec:fullCGCbasics}
In this section, we discuss the derivation of the dipole amplitude from the Color Glass Condensate effective theory \cite{McLerran:1993ni,McLerran:1993ka,McLerran:1994vd,Gelis:2010nm,Albacete:2014fwa}.  For an initial  $x_\mathbb{P}$, it is computed as a stochastic average from fundamental Wilson lines $U(\mathbf{x})$, which are obtained by solving classical Yang-Mills equations with target color sources sampled from 
a local Gaussian distribution
\begin{align}\label{eq:MVgaussian}
\langle  \rho^a(x^-,\mathbf{x}) \rho^b(z^-,\mathbf{z}) \rangle = (g\mu)^2 \delta^{ab}\delta^{2}(\mathbf{x}-\mathbf{z})\delta(x^--z^-)\,,
\end{align}
where the color charge density is related to the IP-Sat value for the saturation scale at moderately large $x_\mathbb{P}$, 
\begin{align}\label{eq:impactparam}
Q_s(\mathbf{x})= c\, g^2\mu (\mathbf{x})\,.
\end{align} 
An impact parameter independent target allows to compute the value of the parameter $c$ directly~\cite{Lappi:2007ku}, and we will vary the parameter between $0.75-0.85$ in the impact parameter dependent case.

For a given target color configuration, the solutions of the Yang-Mills equation specify Wilson lines,
\begin{align}\label{eq:YangMillsSolution}
U(\mathbf{x})= \mathcal{P} \exp\left( -ig \int \der x^- \,\frac{\rho(x^-,\mathbf{x})}{\nabla^2+\tilde{m}^2}\right)\,.
\end{align} 
Here, an IR cutoff $\tilde{m}\sim 0.2-0.4$  GeV has been introduced to avoid nonphysical Coulomb tails at distances $\sim\Lambda_{QCD}^{-1}$ where
non-perturbative effects become important.

After computing Wilson lines at some initial $x_\mathbb{P}$, we determine the energy evolution of the dipole amplitude
using the JIMWLK equations \cite{JalilianMarian:1996xn,JalilianMarian:1997jx,JalilianMarian:1997gr,Iancu:2001md,Ferreiro:2001qy,Iancu:2001ad,Iancu:2000hn}. At leading logarithmic order, the JIMWLK renormalization group equations can be written as
a functional Fokker-Planck equation \cite{Weigert:2000gi} and explicitly expressed as a Langevin equation for the stochastic Wilson lines $U(\mathbf{x})$,
\begin{align}\label{eq:defJIMWLK}
\frac{\der U(\mathbf{x})}{\der y} = U(\mathbf{x})(it^a)\left\{ \int \der^2 z\, \epsilon^{ab,i}(\mathbf{x},\mathbf{z})\, \xi^b_i(\mathbf{z},y) + \sigma^a(\mathbf{x})\right\}\,,
\end{align}
where $\der y=\der x_\mathbb{P}/x_\mathbb{P}$ is the rapidity and $t^a$ are  $SU(N_c)$ generators in the fundamental representation. The drift term in \Eq{eq:defJIMWLK}  reads
\begin{align}\label{eq:difttermJIMWLK}
\sigma^a(\mathbf{x})=-i\frac{\alpha_s}{2\pi^2}\int \der^2 z\, \frac{1}{(\mathbf{x}-\mathbf{z})^2} \text{Tr}[T^a \tilde{U}^\dagger (\mathbf{x}) \tilde{U}(\mathbf{z})  ]\,,
\end{align}
where the Wilson lines $\tilde{U}(\mathbf{z})$, $ \tilde{U}^\dagger(\mathbf{x})$ and generators $T^a$ are in the adjoint representation. The $\xi^b_i(\mathbf{z},y)$'s are stochastic and are sampled from
a Gaussian distribution with zero mean and variance given by
\begin{align}
\langle \xi^a_i(\mathbf{z},y) \xi^b_j(\mathbf{z}',y') \rangle = \delta^{ab}\delta^{ij}\,\delta^{(2)}(\mathbf{z}-\mathbf{z}')\,\delta(y-y')\,.
\end{align}
The kernel in \Eq{eq:defJIMWLK} is
\begin{align}
 \epsilon^{ab,i}(\mathbf{x},\mathbf{z})=\sqrt{\frac{\alpha_s}{\pi}} K^i [1-\tilde{U}^\dagger (\mathbf{x}) \tilde{U}(\mathbf{z}) ]^{ab}\,,
\end{align}
where we abbreviated $K^i\equiv  {(x^i-z^i)}/{(\mathbf{x}-\mathbf{z})^2} $. It is possible to eliminate the drift term by writing \cite{Lappi:2012vw}
\begin{align}
U(\mathbf{x},y+\der y)=\exp\big\{-i\frac{\alpha_s \der y}{\pi} \int \der^2z\, \mathbf{K}\cdot (U \boldsymbol{\xi}U^\dagger )(\mathbf{z}) \big\}\nonumber\\
\times U(\mathbf{x},y) \exp\big\{i\frac{\alpha_s \der y}{\pi} \int \der^2z\, \mathbf{K}\cdot \boldsymbol{\xi}(\mathbf{z}) \big\}\,.
\end{align}
Within the CGC JIMWLK evolution, long distance tails are encountered in the Coulomb kernel $K^i$.  These lead to an exponential growth of the cross section
with rapidity \cite{Kovner:2001bh,GolecBiernat:2003ym,Berger:2010sh}, ultimately violating the Froissart unitarity bound \cite{Froissart:1961ux,Martin:1962rt} unless regulated by non-perturbative physics at large distance scale.  
To regularize the non-perturbative regime, we follow the prescription of \cite{Schlichting:2014ipa} and instead use the kernel
\begin{align}\label{eq:JIMWLKkernel}
\tilde{K}^i(\mathbf{x})\equiv m |\mathbf{x}|K_1(m|\mathbf{x}|)\frac{x^i}{\mathbf{x}}\,,
\end{align}
where $m\sim \Lambda_{QCD}$ and $K_1$ is a modified Bessel function.

We include running coupling effects in our analysis and evaluate the coupling constant as follows \cite{Lappi:2012vw},
\begin{align}\label{eq:runningcoupling}
\alpha_s(r)=\frac{12\pi}{(11N_c-3N_f)\log\left[ \left(\frac{\mu_0^2}{\Lambda_{QCD}^2}\right)^{\frac{1}{\chi}} + \left(\frac{4}{r^2\Lambda_{QCD}^2}\right)^{\frac{1}{\chi}}\right]^\chi}\,,
\end{align}
where $r\equiv |\mathbf{x}-\mathbf{z}|$, $\mu_0=0.28$ GeV and $\chi=0.2$ \cite{Mantysaari:2018zdd}.  In practice, we fix the initial condition for the JIMWLK evolution at $x_\mathbb{P}=0.01$ from the IP-Sat model.
We then evolve towards smaller $x_\mathbb{P}$ by solving the JIMWLK evolution equations. We study different choices of IR regulators $\tilde{m}$ and $m$, in a range which is constrained by and consistent with data \cite{Mantysaari:2016jaz,Mantysaari:2018zdd},
and we compare fixed versus running coupling $\alpha_s$, c.f. \Eq{eq:runningcoupling}. For more details, we refer the reader to \cite{Mantysaari:2018zdd}.
%
%
%
% -------------------  Gluon Wigner at small $x$ distributions from the Color Glass Condensate -------------------------------------------------------------------------------
%
%
%
%
% -------------------  Gluon Wigner at small $x$ distributions from the Color Glass Condensate -------------------------------------------------------------------------------
%
\section{Gluon Wigner and Husimi distributions at small $x$ from the Color Glass Condensate}\label{sec:WIgnerfromCGC}
In the CGC approach, angular correlations between impact parameter and dipole orientation in the dipole amplitude $\mathcal{N}(\rt,\bt,x)$ emerge naturally and need not be modeled, in contrast to e.g. the situation in the IP-Sat model \Eq{eq:IPSatwithcorr}. 

\begin{figure}[tb]
\begin{center}
\includegraphics[width=0.49\textwidth]{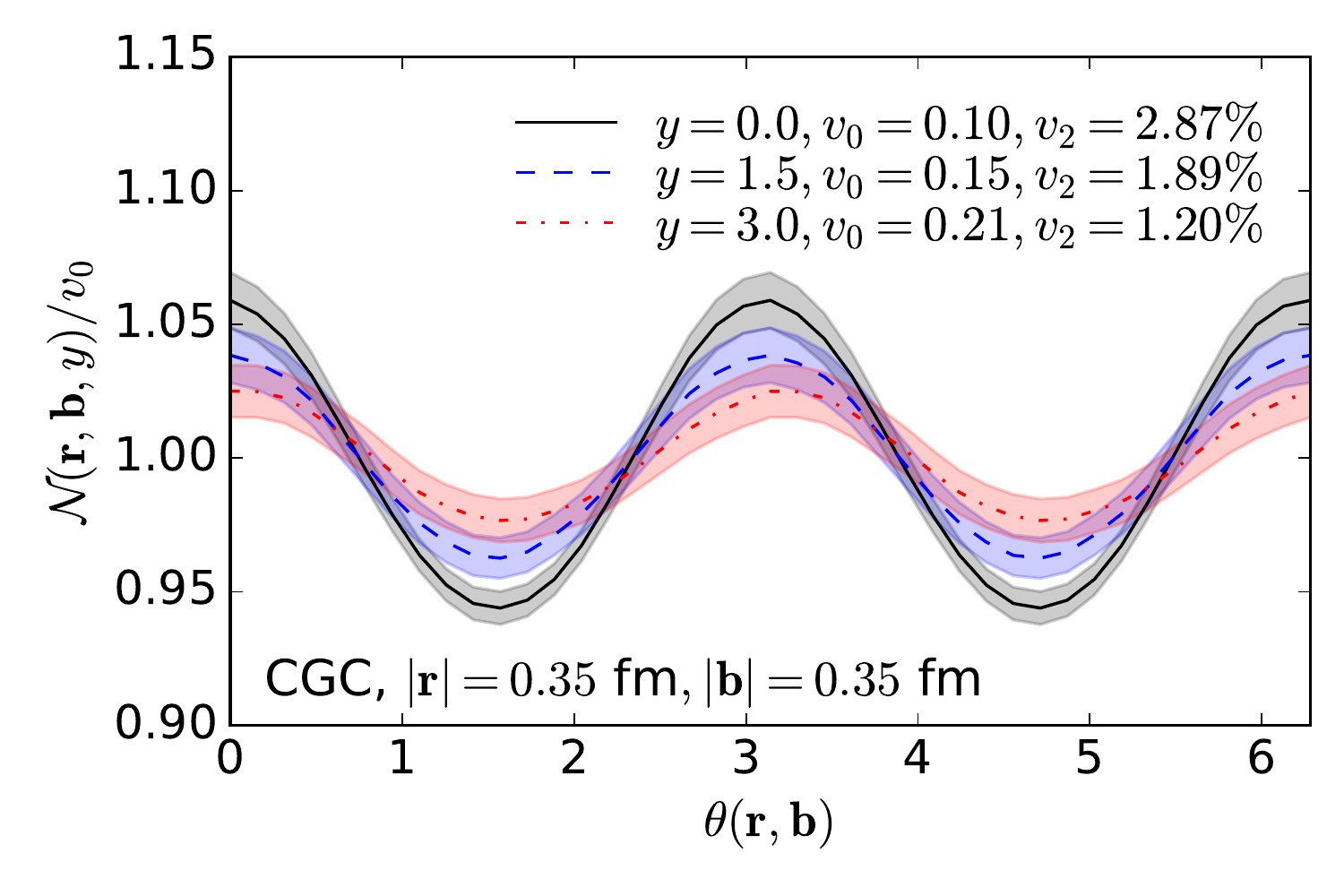}
\end{center}
\caption{The dipole scattering amplitude in the Color Glass Condensate framework as a function of angle $\theta(\rt,\bt)$ between dipole size $\rt$ and impact parameter $\bt$, at $y=0$ and after the JIMWLK evolution up to $y=1.5$ and $y=3.0$. The results are normalized by the average, $v_0$.}
\label{fig:r_b_angle}. 
\end{figure}

In this section, we present a CGC study with fixed $\as=0.21$ and infrared regulators $\tilde m = 0.4$ GeV and $m=0.2$ GeV as in~\cite{Mantysaari:2018zdd}, and obtain the dipole
amplitude by solving the JIMWLK evolution equations with rapidity defined as
\begin{align}
x(y) = x(0)\, e^{-y}\,,
\end{align} 
where $x(0)=10^{-2}$. In Fig.~\ref{fig:r_b_angle}, we show the normalized dipole amplitude  as a function of the relative angle $\theta(\rt,\bt)$ between impact parameter $\bt$ and dipole orientation $\rt$. We present
results for the initial condition ($y=0$) and after $y=1.5$ and $y=3$ units of rapidity evolution.  Here, the dipole amplitude is largest whenever the dipole is oriented along the impact parameter, which is expected as this configuration is most sensitive to (radial) color gradients (for a similar analysis using impact parameter dependent BK evolution, see \cite{Berger:2010sh}).  

\begin{figure}[tb]
\begin{center}
\includegraphics[width=0.49\textwidth]{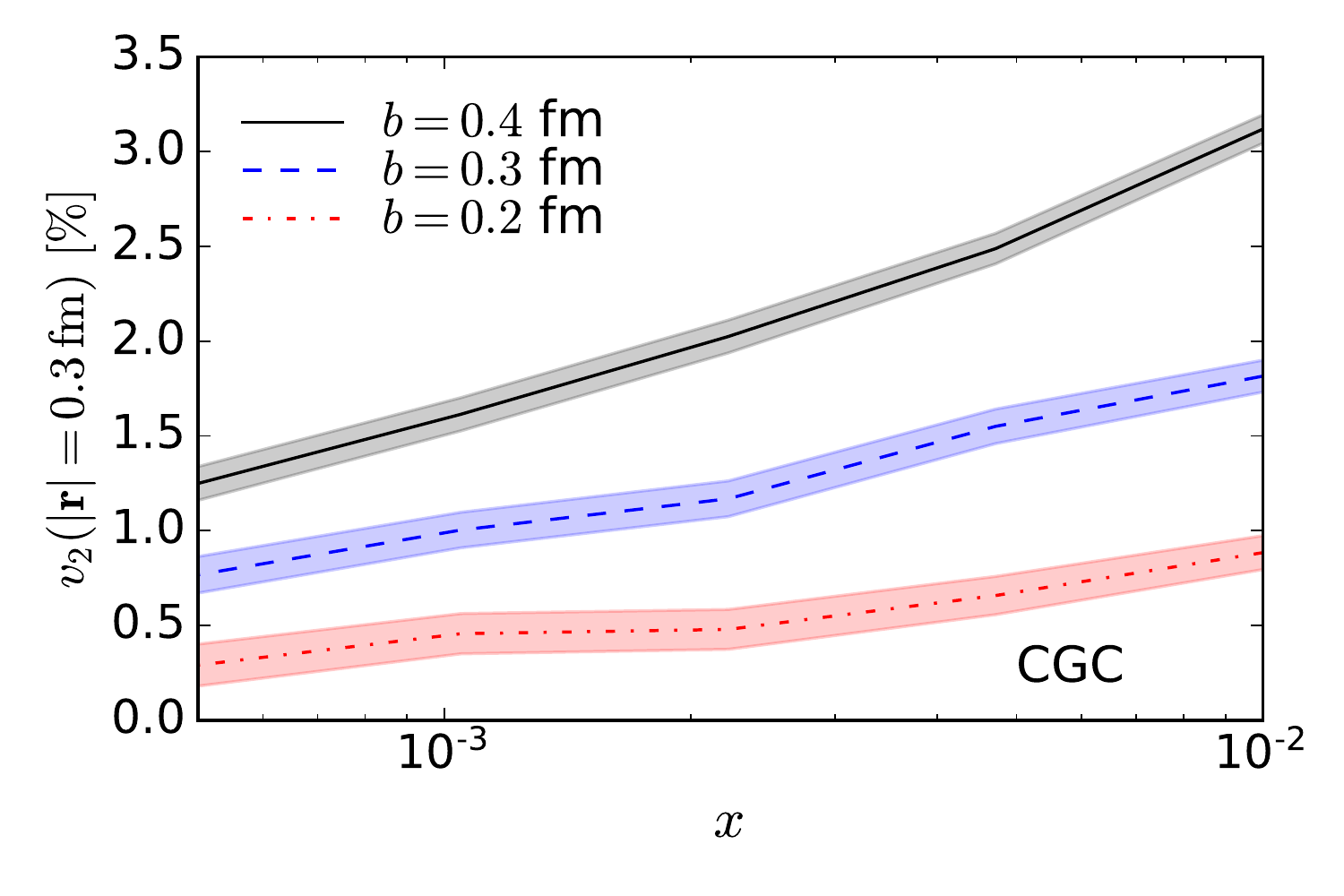}
\end{center}
\caption{Energy (x or rapidity) dependence of the elliptic component of the dipole amplitude $\mathcal{N}(\rt,\bt.x)$ over $3$ units of rapidity for different values of $|\bt|$.
 }\label{fig:dipole_v2_xdep}. 
\end{figure}

To quantify this behavior, the 
elliptic component $v_2$ of the dipole amplitude is defined as
\begin{equation}\label{eq:dipolparametrization}
\mathcal{N}(\rt,\bt,x) = v_0\left[ 1 + 2 v_2 \cos(2\theta(\rt,\bt)) \right],
\end{equation}
where $v_0$ is the average dipole amplitude. We plot the $x$ dependence of $v_2$ for different impact parameters in Fig.~\ref{fig:dipole_v2_xdep} and find that the energy (rapidity or $x$) evolution suppresses the elliptic component significantly.
This is a consequence of the proton's growth with energy, leading to smoother density gradients for fixed impact parameter. We observe a similar trend for the energy dependence
of diffractive dijet cross sections presented in Section \ref{sec:fullCGC}. Further discussion of the proton size dependence can be found in Appendix.~\ref{app:protonsize}.

Next, we study the Wigner and Husimi gluon distributions, which encode information about the transverse momentum and coordinate dependence of the small-$x$ gluons, as discussed in Section ~\ref{sec:WignerIntro}. The quasi-probabilistic gluon Wigner distribution $xW(x,\Pt,\bt)$ is real but not necessarily positive; following~\cite{Hagiwara:2016kam} its small $x$ (CGC-) limit can be computed from \Eq{eq:CGClimitWIgner}, and reads
\begin{multline}
\label{eq:wigner_dipole}
xW(x,\Pt,\bt) = -\frac{2 N_\mathrm{c}}{\alpha_s} \int \frac{\der^2 \rt}{(2\pi)^2} e^{i \Pt \cdot \rt} \\
\times \left(\frac{1}{4} \nabla_\bt^2 + \Pt^2 \right) \mathcal{N}(\rt,\bt, x).
\end{multline}

The Husimi distribution is positive semidefinite and can be computed from the Wigner distribution by Gaussian smearing of transverse momentum and impact parameter.
Following~\cite{Hagiwara:2016kam}, it can be written as
\begin{equation}
\label{eq:husimi_def}
\begin{split}
xH(x,\Pt,\bt) = \frac{1}{\pi^2} \int \der^2 \bt' \der^2 \Pt' e^{-\frac{1}{l^2}(\bt-\bt')^2 - l^2(\Pt-\Pt')^2} \\
\times xW(x,\Pt',\bt')\,,
\end{split}
\end{equation}
where $l$ is an arbitrary smearing parameter. 

We investigated various values for $l$ and present results for $l=1\,\gev^{-1}$ only. This is
 a reasonable choice, ensuring the spatial smearing scale to be smaller than the proton,  while keeping also the (inversely proportional) momentum smearing scale 
small enough to resolve the transverse momentum spectrum of soft gluons in the proton's wave function. However, the dependence on this smearing scale is a disadvantage for interpretations of the Husimi distribution.
We argue below that while Husimi and Wigner distributions agree in certain limits where smearing has no effect, some of the  former's features can be $l$-dependent,
making it more favorable to work with the Wigner distribution instead. 

Similarly as for the dipole amplitude, we study the elliptic modulation of Wigner and Husimi distributions. Because $v_2$ (in \Eq{eq:dipolparametrization}) is not defined at zero crossings of the
respective distribution (recall that the Wigner distribution is not positive definite), we proceed with a different parameterization~\cite{Hagiwara:2016kam}\footnote{In Appendix~\ref{app:detailsWignerCGC} we outline a practical computation of these coefficients.}
\begin{equation}
xW(\Pt,\bt,x) = xW_0 + 2 xW_2 \cos(2\theta(\Pt,\bt))\,,
\end{equation}
and similar for the Husimi distribution,
\begin{equation}
xH(\Pt,\bt,x)  = xH_0 + 2 xH_2 \cos(2 \theta(\Pt,\bt))\,. 
\end{equation}

\begin{figure}[tb]
\begin{center}
\includegraphics[width=0.49\textwidth]{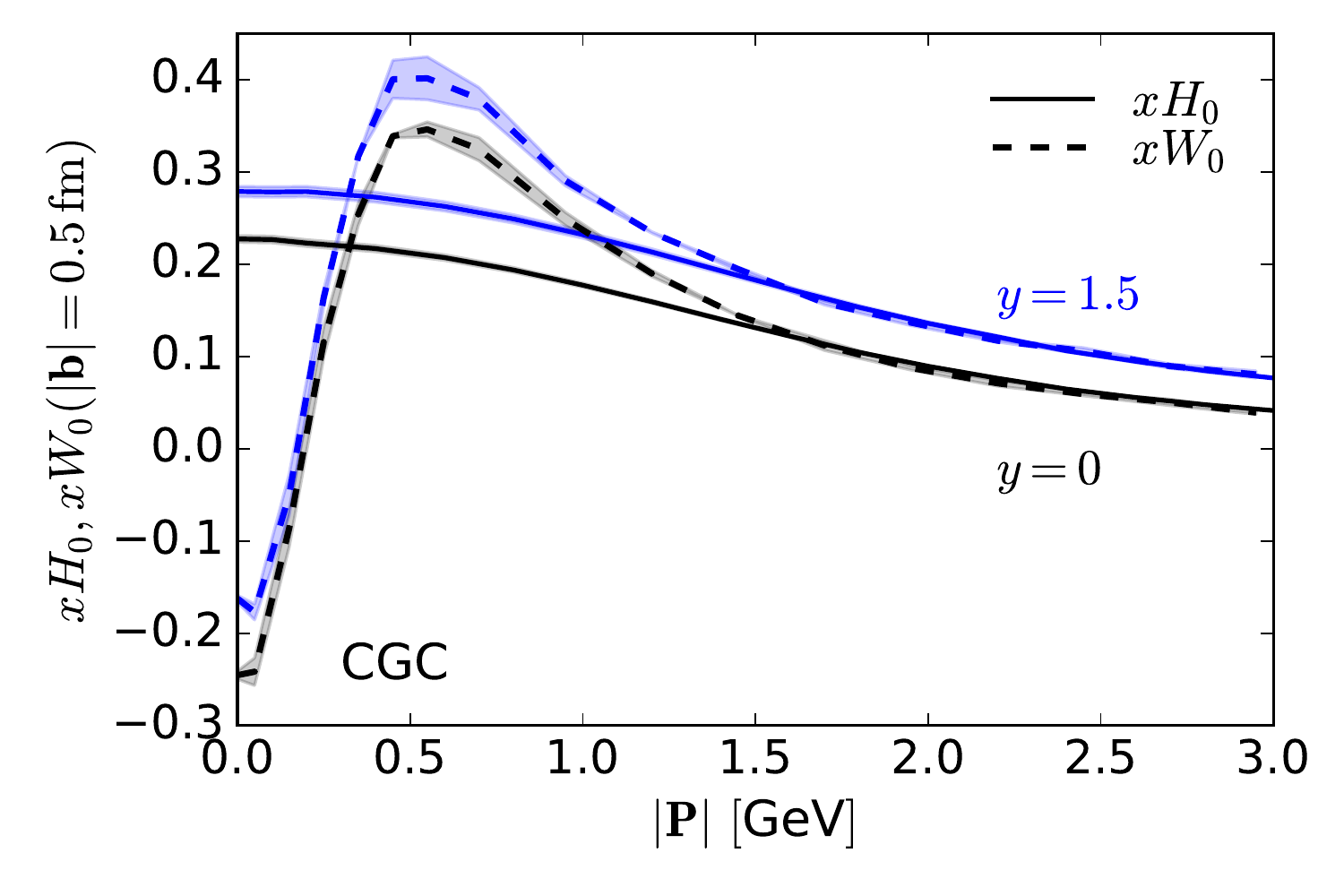}
\end{center}
\caption{Wigner and Husimi distributions from the CGC at $y=0$ (lower black lines) and after $1.5$ units of rapidity evolution (upper blue lines) as a function of transverse momentum $|\mathbf{P}|$. These results are averaged over the azimuthal angle between the transverse momentum and the impact parameter.}
\label{fig:husimi_wigner_v0}. 
\end{figure}

In \Fig{fig:husimi_wigner_v0}, we show the lowest moments $xW_0$ and $xH_0 $ of Wigner and Husimi distributions as a function of transverse momentum and for different
rapidities $y=0$ and $y=1.5$. The effect of smearing is most prominent at small momenta, where $|\mathbf{P}| l \ll 1 $. At large momenta, where both Husimi and Wigner 
distributions are positive, smearing has no effect and the distributions agree.
\begin{figure}[tb]
\begin{center}
\includegraphics[width=0.49\textwidth]{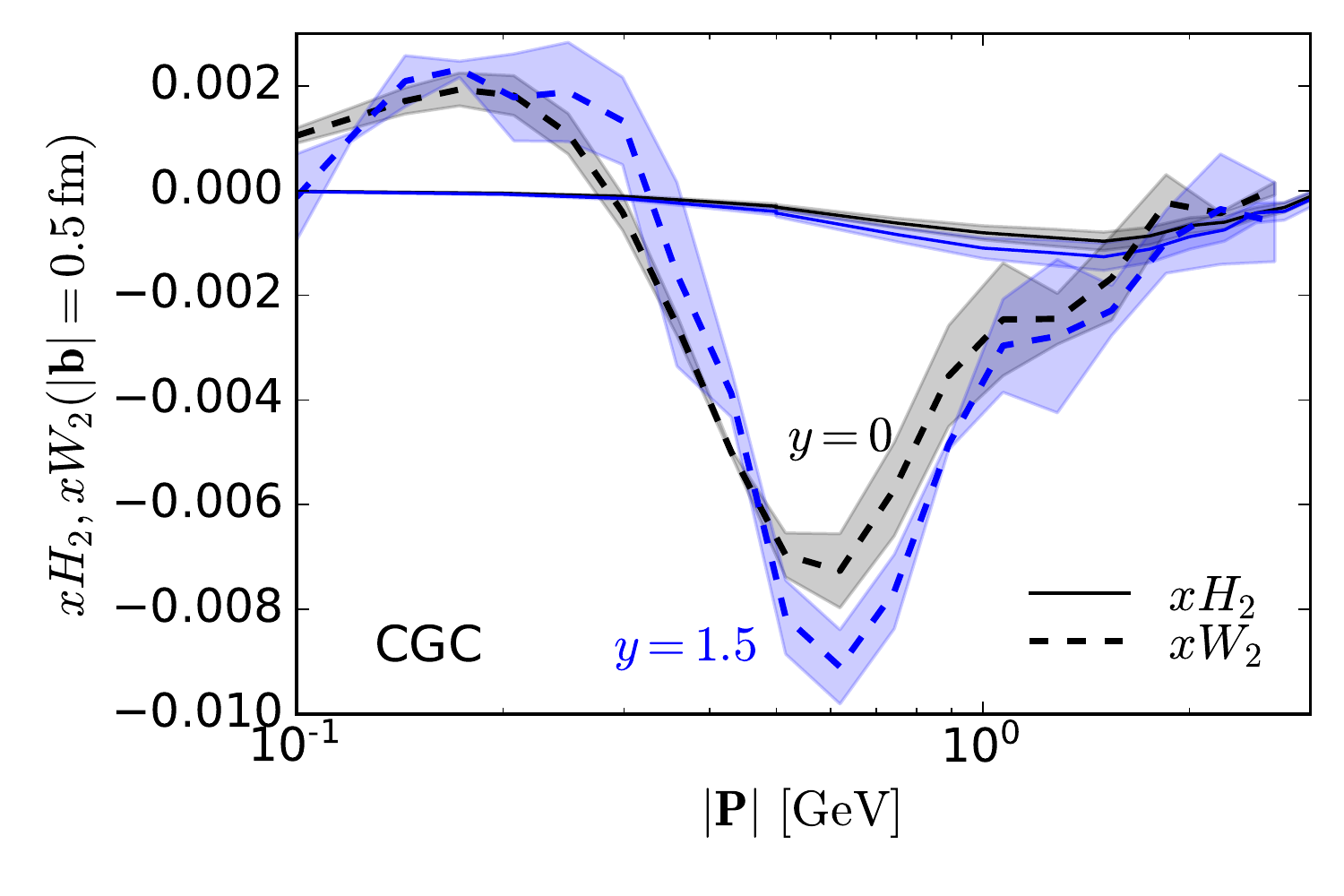}
\end{center}
\caption{Elliptic components $xW_2$ and $xH_2$ of  Wigner and Husimi distributions at $y=0$ (black lines) and after $1.5$ units of rapidity evolution (blue lines), as a function of transverse momentum.}
\label{fig:husimi_wigner_v2}. 
\end{figure}
The presented Wigner distribution should be contrasted with the results of a previous analysis of gluon Wigner distributions~\cite{Hagiwara:2016kam}, with results relatively
similar to ours. Overall we find qualitative agreement for $|\mathbf{P}|\gtrapprox 0.1$ GeV, however we do not find that position of the characteristic peak in \Fig{fig:husimi_wigner_v0} evolves with energy in our study. We also do not recover a feature, seen in~\cite{Hagiwara:2016kam}, where  the Wigner distribution rises from below to zero at very small momentum. 
 
\begin{figure}[tb]
\begin{center}
\includegraphics[width=0.49\textwidth]{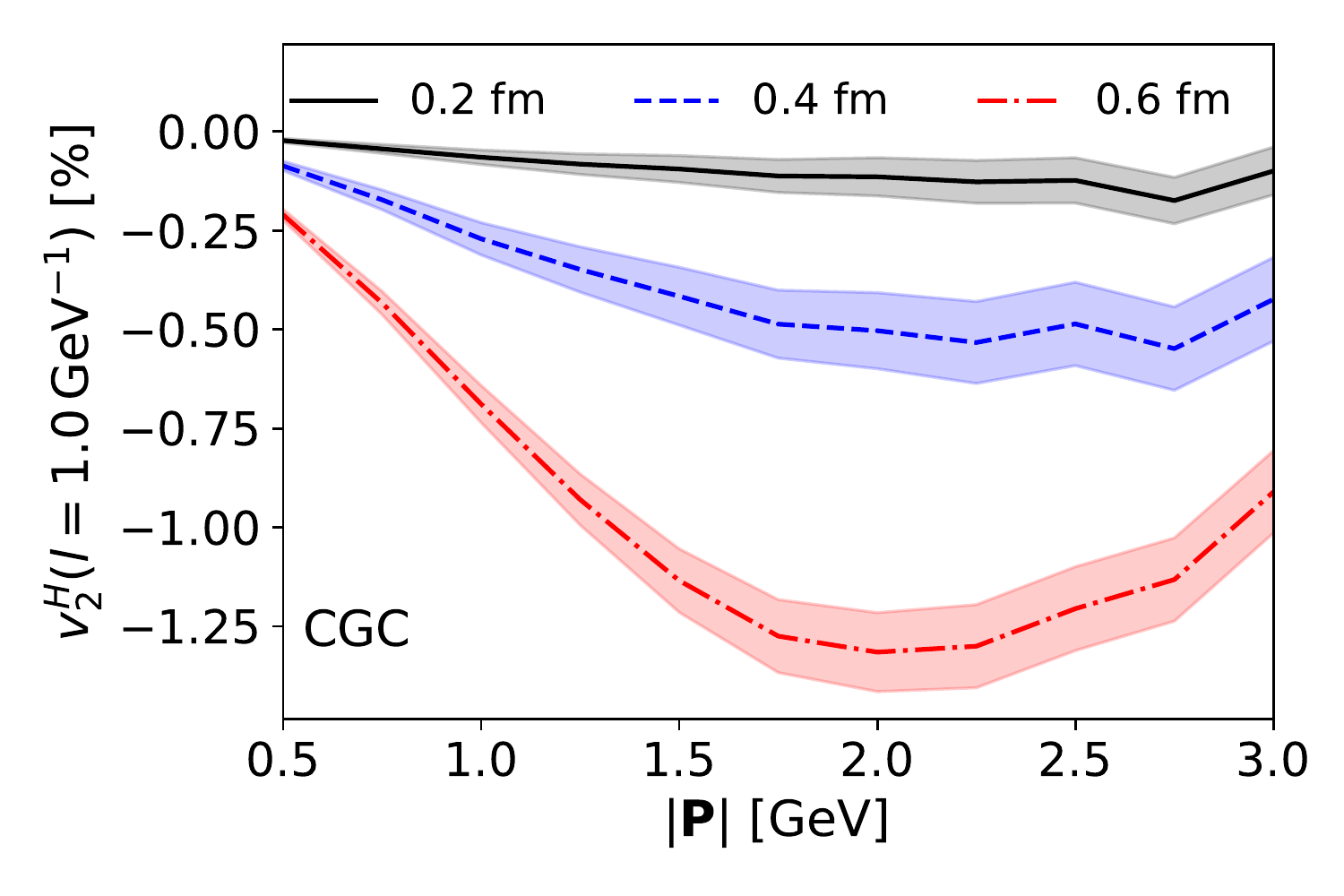}
\end{center}
\caption{Elliptic coefficient $v_2^H$ of the Husimi distribution at $y=0.0$ for different impact parameters, shown as a function of transverse momentum $|\mathbf{P}|$.}
\label{fig:husimi_b_P}. 
\end{figure}
We are not surprised by the differences between the model study~\cite{Hagiwara:2016kam} and our numerical CGC computation. At small momentum, (nonperturbatively-) large dipoles (as large as a few fm) give a significant contribution
to the Wigner distribution, while the only effective regulator of these contributions is the proton size. In our framework this extreme IR behavior is parameterized by the infrared regulators $m$ and $\tilde{m}$  (see Eqs.~\eqref{eq:YangMillsSolution} and \eqref{eq:JIMWLKkernel}), which are constrained to some degree by HERA data\footnote{A more detailed discussion of the IR regulator dependence can be found in Section \ref{sec:dijetCS}.}. In contrast, the authors of~\cite{Hagiwara:2016kam} introduced an explicit exponential suppression factor (by hand) to cut off contributions from this regime\footnote{Note that large distance contributions to the Husimi distribution are suppressed by the smearing scale $l$.}.
 
 In \Fig{fig:husimi_wigner_v2} we show the elliptic components $xW_2$ and $xH_2$ as a function of transverse momentum. We find the magnitude of $xW_2$
to be on the percent level when compared to $xW_0$, similar as in~\cite{Hagiwara:2016kam}. We observe that the elliptic part of the Wigner distribution is much larger than that of the Husimi distribution
 for $|\mathbf{P}| \lesssim 2$ GeV. It is numerically difficult to extract $xH_2$ at larger momentum, but we find that the distributions agree in this regime within error bands shown
 in \Fig{fig:husimi_wigner_v2}. The small $|\mathbf{P}|$ behavior of  $xW_2$ and $xH_2$ is again sensitive to infrared regularization and deserves further study.

In order to gain intuition and to illustrate some of the difficulties in its interpretation, we study the Husimi distribution in more detail. For better comparison with the $v_2$ component of the dipole amplitude and of the dijet cross section studied below, we present the normalized elliptic coefficient
in \Fig{fig:husimi_b_P},
\begin{equation}
v_2^H = \frac{xH_2}{ xH_0 }\,.
\end{equation}
Here,  we show the momentum dependence of $v_2^H$ for $y=0$ ($x=10^{-2}$) at different impact parameters. As expected,   $v_2^H$ vanishes at small $|\bt|$ where the average proton color profile is nearly 
homogeneous and cannot have any angular dependence. The elliptic component peaks at a characteristic moderately large $|\Pt|$ (the exact position of the peak depends on the parameter $l$) and approaches zero at small and large momentum. 
\begin{figure}[tb]
\begin{center}
\includegraphics[width=0.49\textwidth]{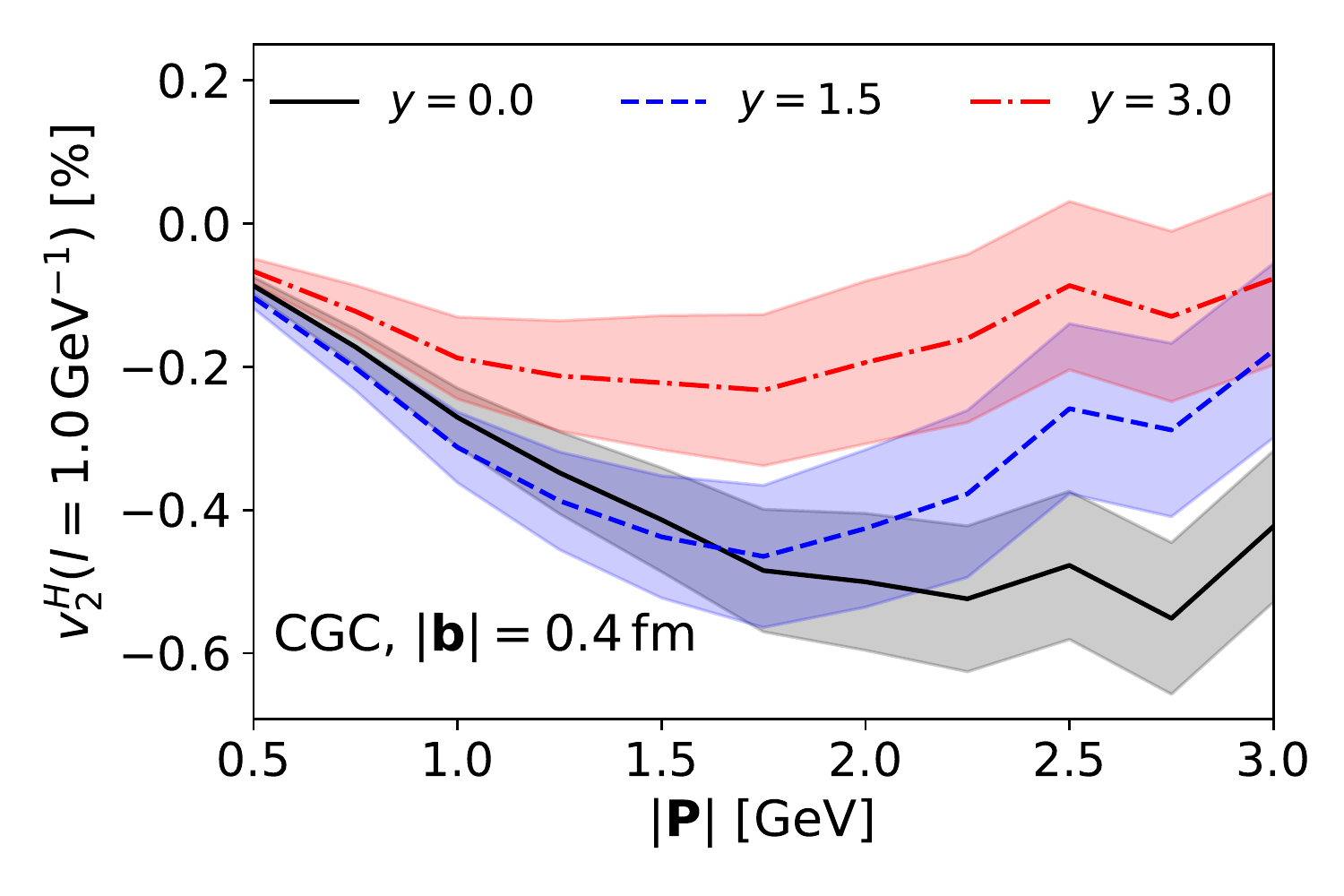}
\end{center}
\caption{Rapidity evolution of the elliptic part of the Husimi distribution as a function of transverse momentum $|\mathbf{P}|$.}
\label{fig:husimi_b_P_evolution}. 
\end{figure}

In \Fig{fig:husimi_b_P_evolution}, we show the energy (rapidity) dependence of $v_2^H$ at fixed impact parameter, where again the growth of the proton with energy leads to a decrease of $v_2^H$,
 which may be compared to the $v_2$ component of the dipole amplitude and of the dijet cross section studied below. Remarkably, the decrease of $|v_2^H|$ with energy, caused by the decreasing density gradients, is not uniform for all $|\mathbf{P}|$.
 In the small momentum regime, the decrease of $|v_2^H|$ is prevented by the fact that when the proton grows, one first starts to include dipoles with transverse separation $|\rt|\sim |\mathbf{P}|^{-1}$ with large elliptic modulation. Eventually, the proton grows larger than these dipoles and the probed density gradients become smaller, resulting in a (delayed) decrease of $|v_2|$ at small transverse momentum. 
 
This behavior exemplifies some of the difficulties in assigning a physical interpretation to the Husimi distribution. For example, with smaller $l$ (less smearing in transverse coordinate space), the dipole sizes would be more strongly limited and $|v_2^H|$ would uniformly decrease for all $|\mathbf{P}|$.\footnote{One could optimize the choice of $l$ for each $y$, thereby finding the optimal resolution for given proton size and characteristic transverse momentum scale $Q_s$. In this way one could reduce the parameter dependence of the elliptic modulation in the Husimi distribution, but this is not a very attractive option.}
While the positivity of the distribution might be an appealing argument for a probabilistic interpretation, we wish to argue that more robust information can be extracted from the Wigner distribution.
 
 In the next section we compute diffractive coherent dijet production cross sections in electron-proton scattering at typical EIC energies. Here, we study
 angular correlations between the dijet transverse momentum and proton recoil, which are directly sensitive to correlations between impact parameter
 and transverse momentum in the gluon Wigner and Husimi distributions. 
\begin{figure}[tb]
\includegraphics[width=0.49\textwidth]{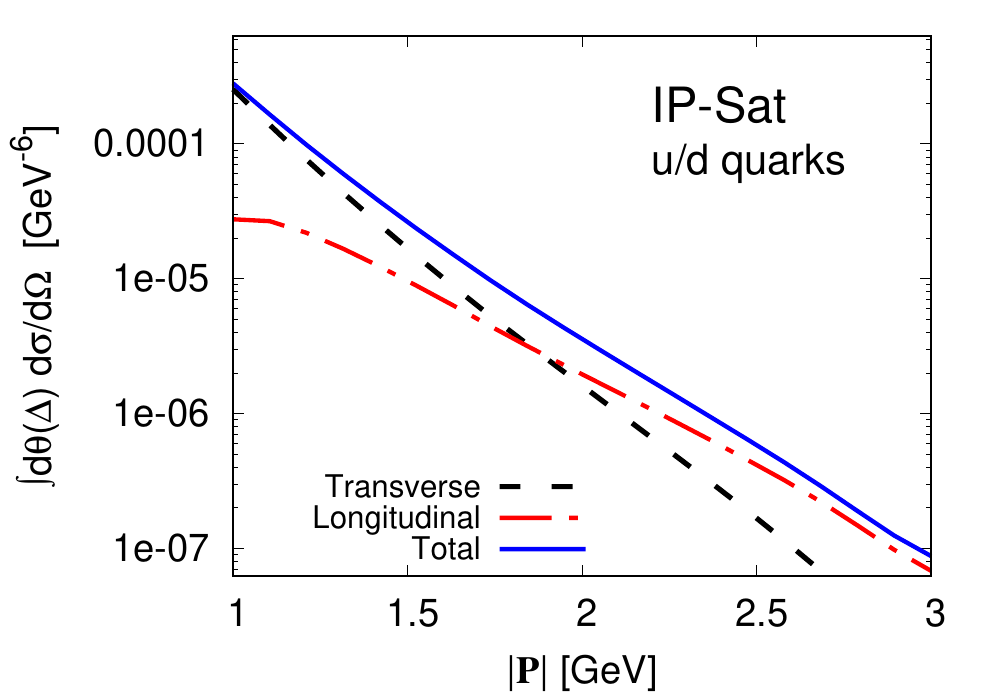}
\caption{Angle integrated transverse (dashed), longitudinal (dash-dotted) and total cross section (solid) for dijet with either up or down quarks, as a function of $|\mathbf{P}|$, where $|\mathbf{\Delta}|=0.1$ GeV and $z=\bar{z}=0.5$. }
\label{fig:IPSatMagnitude}
\end{figure}
\begin{figure}[tb]
\includegraphics[width=0.49\textwidth]{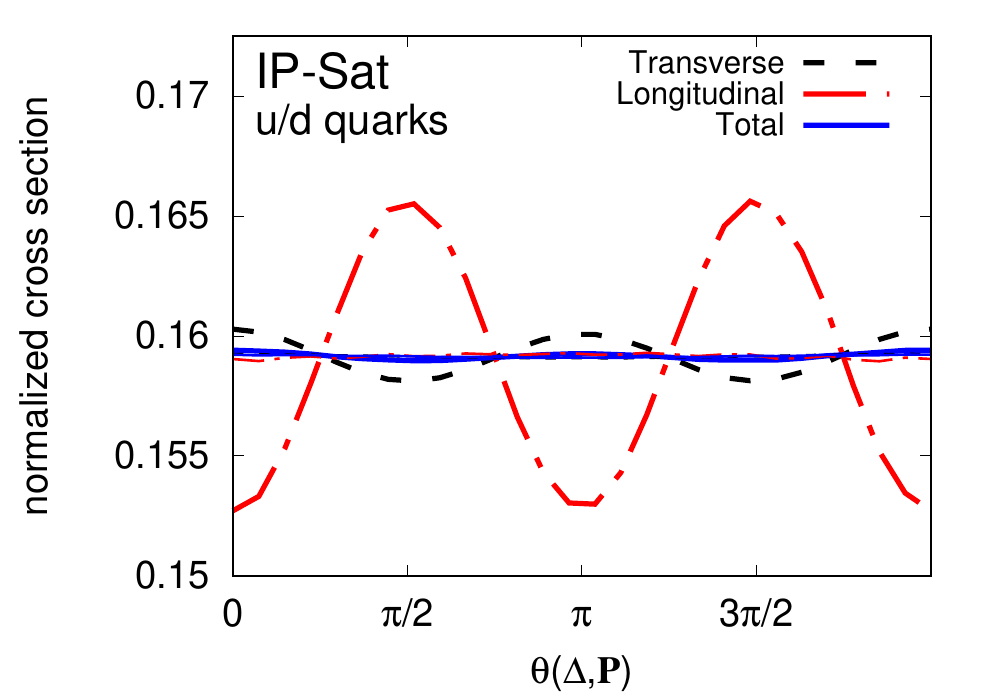}
\caption{Normalized transverse (dashed), longitudinal (dash-dotted) and total cross sections (solid) for light quarks as a function of the relative angle $\theta(\mathbf{\Delta},\mathbf{P})$, from the IP-Sat dipole 
with (thick lines) and without (thin lines) angular correlation between $\mathbf{r}$ and $\mathbf{b}$, c.f. \Eq{eq:IPSatwithcorr}. Here $|\mathbf{P}| = 1$ GeV, $|\mathbf{\Delta}| =0.1$ GeV, $Q^2=1$ $\text{GeV}^2$. The cross section
is integrated over $z\in[0.1,0.9]$ and over the direction of the proton recoil $\theta(\mathbf{\Delta})$.}
\label{fig:IPSatoverviewnormalized}
\end{figure}
% ---------------------------------------------------------------------------------------------------------------------------------------------------------------------------------------------------------------------
% ----------------------------- Angular correlations in diffractive dijet production at the Electron Ion Collider ----------------------------------------------------------------------------------
% ---------------------------------------------------------------------------------------------------------------------------------------------------------------------------------------------------------------------
%
\section{Coherent diffractive dijet production}\label{sec:dijetCS}
In this section, we present results for coherent diffractive dijet production in virtual photon-proton scattering. We investigate angular correlations between transverse dijet momenta and target recoil.  
We focus on typical EIC kinematics and for most of our study we fix the proton beam energy $E_p=250$ GeV, and the center of mass energy to be $W=\sqrt{(P+q)^2}=100$ GeV. Our results are presented in the analysis frame, where the photon and incoming nucleon have no transverse momentum and our convention is that the photon (proton) has a large $+$ ($-$) momentum. Below, we first discuss results from the IP-Sat parameterization, which we use to disentangle kinematic effects from genuine correlations in the underlying gluon dipole distribution.
%
% ---------- baseline IPSat ------------------------------------------------------------------------------------------------------------------------------------------------------------------------------------
%
\subsection{Baseline study: Angular correlation in the modified IP-Sat model}\label{sec:baselineIPSat}

\begin{figure}[tb]
\includegraphics[width=0.49\textwidth]{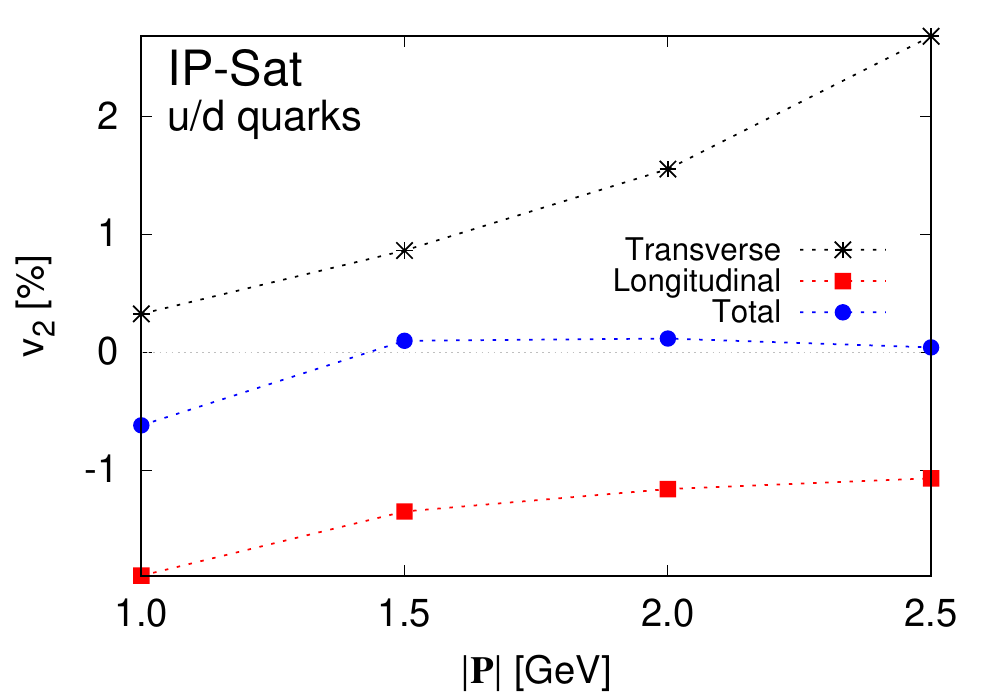}
\caption{Elliptic Fourier coefficient $v_2$ for light quark dijets as a function of $|\mathbf{P}|$, integrated over $z\in [0.1,0.9]$, for the IP-Sat model with angular correlations. Here, $|\mathbf{\Delta}| =0.1$ GeV, $Q^2=1$ $\text{GeV}^2$.}\label{fig:IPSatv2Pdep}
\end{figure}

We begin with an overview of our results for coherent diffractive dijet production from the IP-Sat parameterization, with (\Eq{eq:IPSatwithcorr}) and without (\Eq{eq:IPSat}) angular correlations between impact parameter
and dipole orientation. We present results for typical EIC kinematics, where $E_p=250$ GeV, $W=100 $ GeV and $Q^2=1$ GeV${}^2$. 
\begin{figure*}[tb]
\includegraphics[width=0.89\textwidth]{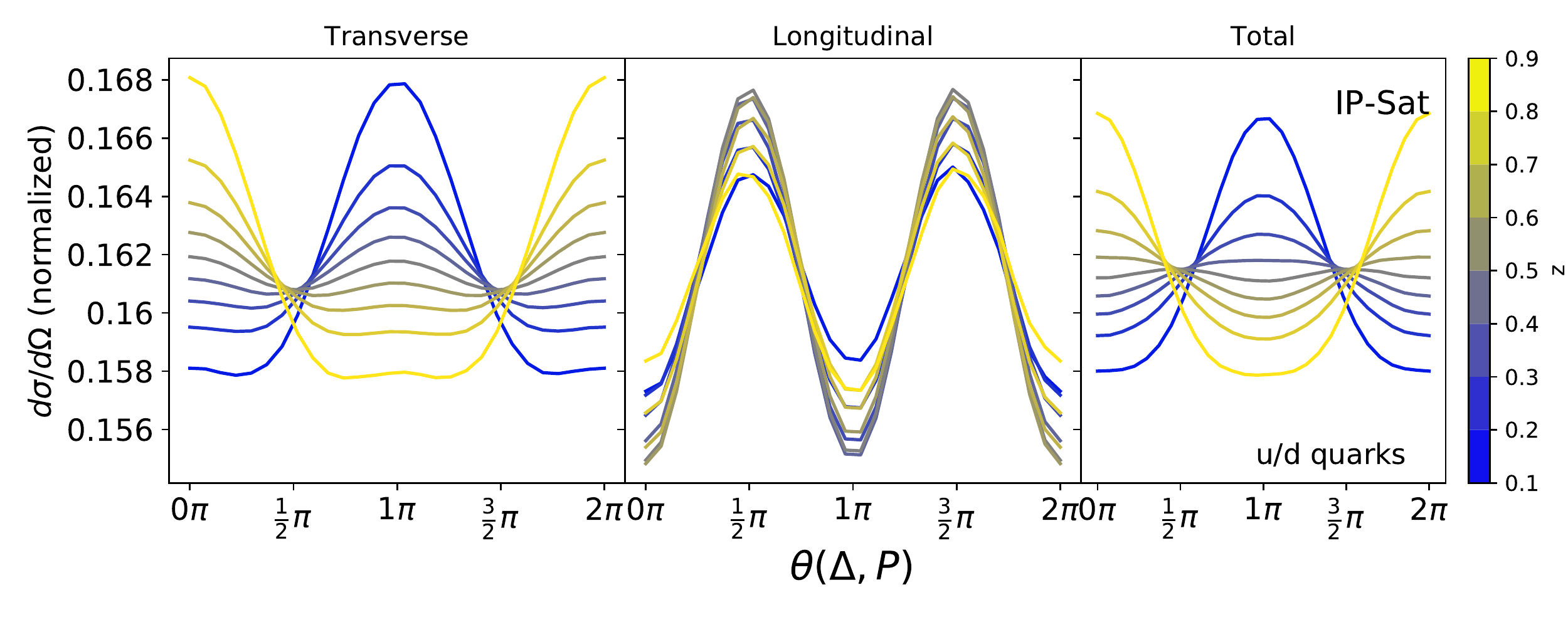}
\caption{Cross section for transversely (left), longitudinally (center) polarized photons and total cross section (right), as a function of the longitudinal momentum fraction of one jet $z=p_0^+/q^+$. A sizable
$v_1$ is induced by kinematic effects, because $x_\mathbb{P}$ is not constant along $\theta(\mathbf{\Delta},\mathbf{P})$ for fixed $z$, according to \Eq{eq:xPomeron}. This effect is $O(|\mathbf{\Delta}|/|\mathbf{P}|)$
and vanishes only in the asymptotic correlation limit.}\label{fig:IPSatzdep}
\end{figure*}

In \Fig{fig:IPSatMagnitude} we show the cross
section for longitudinally and transversely polarized photons as a function of transverse dijet momentum $|\mathbf{P}|$, where we have set $|\mathbf{\Delta}|=0.1$ GeV and 
$\theta(\mathbf{\Delta},\mathbf{P})=\pi$. These results are for light flavor jets with quark mass $m_q=0.03$ GeV and symmetric longitudinal jet momenta $z=\bar{z}=0.5$. As expected, the cross section is steeply falling with increasing jet momentum $|\mathbf{P}|$, and the total
cross section is dominated by the transversely polarized photons at small $|\mathbf{P}|$ and by longitudinal photons at larger $|\mathbf{P}|$.

In \Fig{fig:IPSatoverviewnormalized}, we show normalized dijet cross sections as a function of the relative angle $\theta(\mathbf{\Delta},\mathbf{P})$ between jet transverse momentum $\mathbf{P}$ and target recoil $\mathbf{\Delta}$. Here, we integrate over  longitudinal momentum fraction $z,\bar{z}\in[0.1,0.9]$. We compare results from the conventional IP-Sat parameterization, \Eq{eq:IPSat}, without angular correlations between impact parameter and dipole orientation (thin lines), with those obtained from the modified IP-Sat model \Eq{eq:IPSatwithcorr}, including angular correlations (thick lines). Here, we have set $\tilde{c}=1$, which induces a non-zero dijet $v_2$ of the order of a few percent. The elliptic correlation has opposite sign for longitudinal ($v_{2,L}<0$) and transverse photons ($v_{2,T}>0$).

In \Fig{fig:IPSatv2Pdep} we present longitudinal, transverse and total $v_2$ as a function of the dijet momentum $|\mathbf{P}|$ with kinematics as in \Fig{fig:IPSatoverviewnormalized}. For all presented $|\mathbf{P}|$, longitudinal and transverse components have opposite sign and the total $v_2$ is partially cancelled between the two contributions.
 
\begin{figure}[tb]
\includegraphics[width=0.49\textwidth]{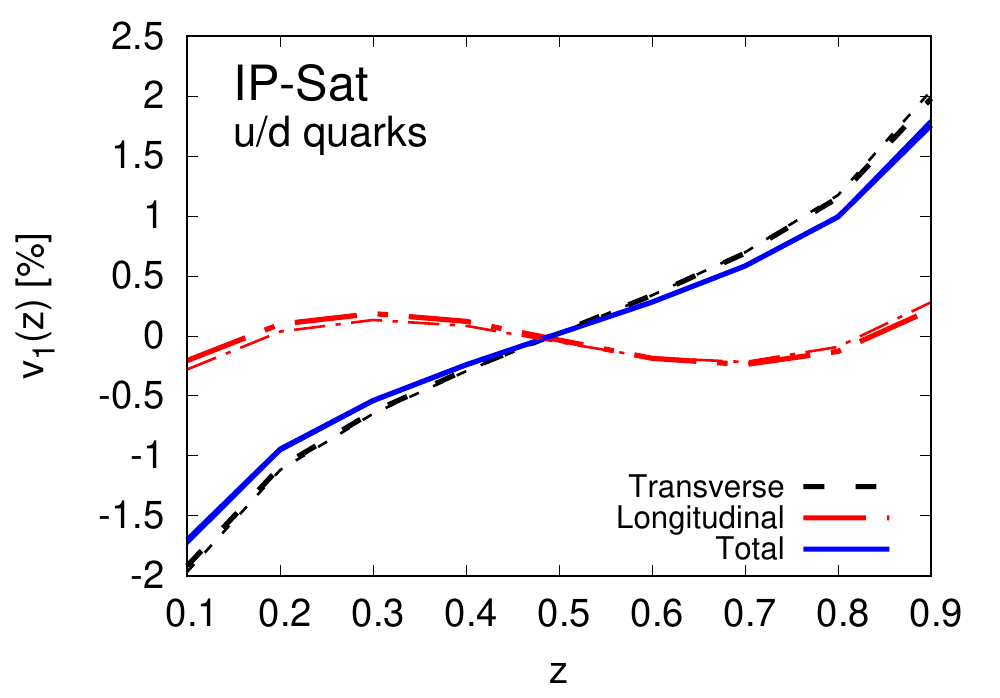}
\caption{Azimuthal Fourier coefficient $v_1=\langle \cos(\theta(\mathbf{P},\mathbf{\Delta})) \rangle$ for light quark dijets as a function of $z$ for the IP-Sat model with (thick lines) and without (thin lines) angular correlation. Here, $|\mathbf{P}| = 1$ GeV, $|\mathbf{\Delta}| =0.1$ GeV, $Q^2=1$ $\text{GeV}^2$. 
}\label{fig:IPSatv1}
\end{figure}

\begin{figure}[tb]
\includegraphics[width=0.49\textwidth]{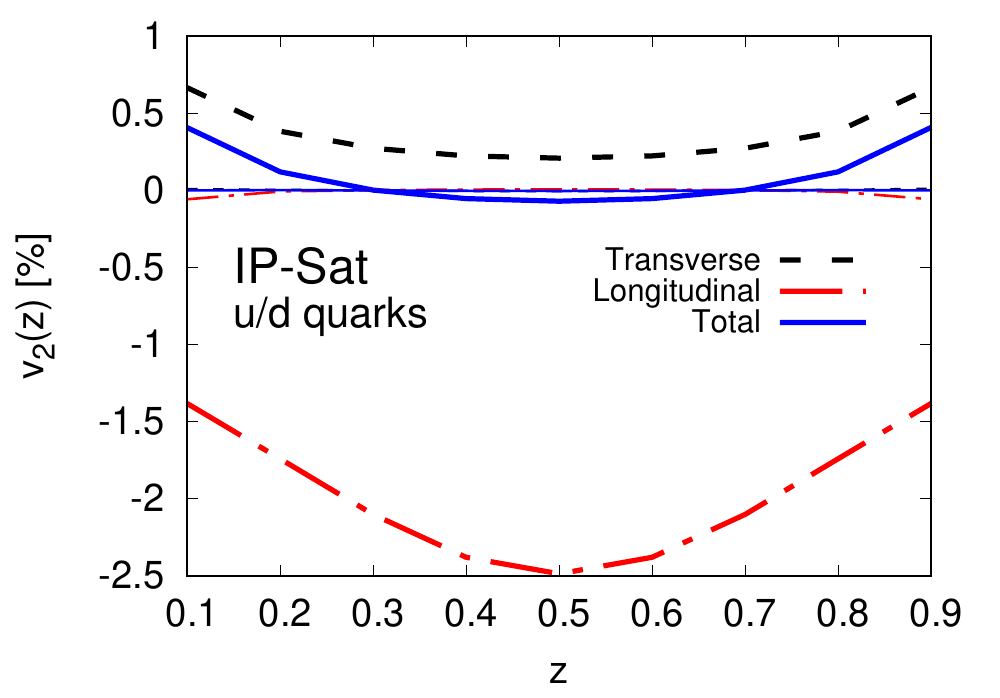}
\caption{Elliptic Fourier coefficient $v_2$ for light quark dijets as a function of $z$ for the IP-Sat model with (thick lines) and without (thin lines) angular correlation. Here, $|\mathbf{P}| = 1$ GeV, $|\mathbf{\Delta}| =0.1$ GeV, $Q^2=1$ $\text{GeV}^2$.}\label{fig:IPSatv2}
\end{figure}

\Fig{fig:IPSatzdep} shows normalized transverse (left), longitudinal (center) and total (right) dijet cross sections as a function of $\theta(\mathbf{\Delta},\mathbf{P})$, separately for different values of $z$. Remarkably, a non-zero $v_1(z)\equiv \langle \cos[\theta(\mathbf{\Delta},\mathbf{P})] \rangle_z$ is observed (most visible in the left panel) whose origin we discuss below.
To illustrate this feature, we plot $v_1(z)$ and $v_2(z)$ in \Fig{fig:IPSatv1} and \Fig{fig:IPSatv2} for transversely (dashed) and longitudinally polarized (dash-dotted) photons with (thick lines) and without (thin lines) correlations between impact parameter and dipole orientation.  The non-zero $v_1(z)$ is independent of the angular correlation in \Eq{eq:IPSatwithcorr}. Its emergence is due to the energy dependence of the dipole amplitude (\Eq{eq:xPomeron}). To show this, we Taylor-expand the dipole amplitude, close to but away from the asymptotic correlation limit to linear order in $|\mathbf{\Delta}|/|\mathbf{P}|$,
 \begin{align}\label{eq:v1contribution}
 \mathcal{N}(\mathbf{r},\mathbf{b},x_\mathbb{P})&\approx \mathcal{N}(\mathbf{r},\mathbf{b},x_\mathbb{P}^0) + \frac{\partial \mathcal{N}(\mathbf{r},\mathbf{b},x_\mathbb{P})}{\partial({|\mathbf{\Delta}|}/{|\mathbf{P}|})}\Big|_{\frac{|\mathbf{\Delta}|}{|\mathbf{P}|}=0} \frac{|\mathbf{\Delta}|}{|\mathbf{P}|}\nonumber\\
&=\mathcal{N}(\mathbf{r},\mathbf{b},x_\mathbb{P}^0) + \frac{\partial \mathcal{N}}{ \partial x_\mathbb{P}} \frac{\partial x_\mathbb{P}}{\partial({|\mathbf{\Delta}|}/{|\mathbf{P}|}) }\Big|_{\frac{|\mathbf{\Delta}|}{|\mathbf{P}|}=0} \frac{|\mathbf{\Delta}|}{|\mathbf{P}|}\,.
\end{align}
The linear term can be computed from \Eq{eq:xPomeron}, 
\begin{align}\label{eq:corrv1}
\frac{\partial x_\mathbb{P}}{\partial({|\mathbf{\Delta}|}/{|\mathbf{P}|}) }\Big|_{\frac{|\mathbf{\Delta}|}{|\mathbf{P}|}=0}  = \frac{\bar{z}-z}{z\bar{z}} \, \frac{|\mathbf{P}|^2}{W^2+Q^2-m_N^2}\,\cos[\theta(\mathbf{\Delta},\mathbf{P})]\,.
\end{align}
If we insert \Eq{eq:corrv1} into \Eq{eq:v1contribution} and use it to compute the dijet cross section, the latter must have a non-zero $v_1(z)$ at $O(|\mathbf{\Delta}|/|\mathbf{P}|)$, vanishing only in the exact correlation limit. 
However, because $v_1(z) =-v_1(\bar{z})$, $v_1(z)$ can be eliminated by $z-$symmetric integration, see for example \Fig{fig:IPSatoverviewnormalized} where $z\in[0.1,0.9]$. 
This effect might have practical implications in experiment where finite detector acceptance (in the laboratory frame) might not allow to eliminate $v_1$\footnote{We note that a different choice of dijet and target recoil transverse momenta, $\tilde{\mathbf{P}}\equiv \bar{z}\mathbf{p}_0 - z\mathbf{p}_1$ and $\tilde{\mathbf{\Delta}}=\mathbf{p}_0+\mathbf{p}_1$,  discussed in Appendix \ref{app:defkinematics},
avoids the $\cos[\theta(\mathbf{\Delta},\mathbf{P})]$-dependence \cite{Dumitru:2018kuw}. These variables cannot be interpreted as Fourier conjugates to impact parameter and dipole orientation and other kinematic effects must be considered.}.

Fig.\,\ref{fig:IPSatv2} shows that a finite $v_2$ only appears for $\tilde{c} \neq 0$. For longitudinal polarization $v_2$ is negative for all $z$ and $|v_2|$ is maximal at $z=0.5$. In contrast, for transverse polarization, $v_2$ is positive for all $z$ and maximal in the most asymmetric cases $z\rightarrow 0$ and $z\rightarrow 1$.

Next we compute the dijet cross section for the case of charm quarks. A simple analysis of \Eq{eq:transvCS} and \Eq{eq:longtCS} shows that dipoles with size $|\mathbf{r}|\gtrapprox \Lambda_\text{QCD}^{-1}$ contribute
to the dijet cross section  for light quarks and small photon virtuality of only a few GeV${}^2$. While the IP-Sat dipole parameterization reports this limit as $\mathcal{N}(|\mathbf{r}|\rightarrow \infty)\rightarrow 1$,  
no reliable theoretical description of angular correlations exists in this regime.  Here, non-perturbative hadronic effects must be taken into account.  From now on we focus on charm dijets, which by means of the large mass scale, allows us to predict cross sections for photon virtuality down to $Q^2\sim 0$ without unconstrained contributions from large distances  $\sim \Lambda_\text{QCD}^{-1}$.

\begin{figure}[tb]
\includegraphics[width=0.49\textwidth]{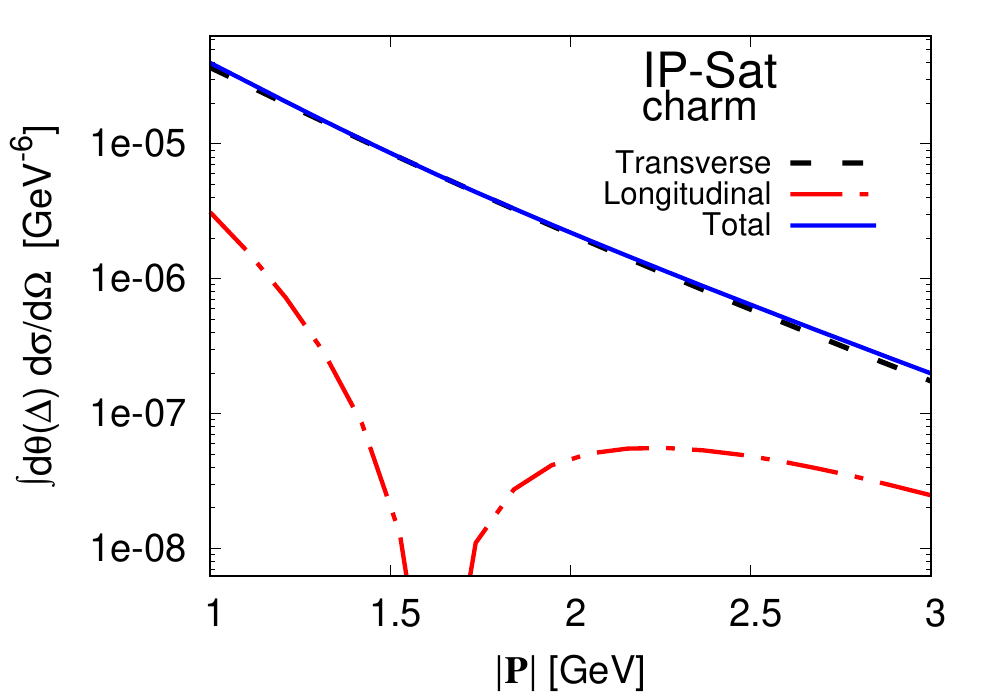}
\caption{$|\mathbf{P}|$-dependence of the angle integrated diffractive dijet production cross section for charm quarks from the modified IP-Sat model for $\tilde{c}=1$. Here, $|\mathbf{\Delta}|=0.1$ GeV, and $z=\bar{z}=0.5$.}\label{fig:IPSat_Pdep_charm}
\end{figure}

Results for the charm dijet cross section from the modified IP-Sat model are shown in \Fig{fig:IPSat_Pdep_charm}. Here, we plot the $|\mathbf{P}|$ dependence of the transverse, longitudinal and total diffractive charm-dijet production cross sections for charm quark mass $m_c=1.28 $ GeV, photon virtuality $Q^2=1$ GeV${}^2$, $|\mathbf{\Delta}|=0.1 $ GeV,  $\theta(\mathbf{\Delta},\mathbf{P})=\pi$ and $z=\bar{z}=0.5$. In the shown range of $|\mathbf{P}|$ the transverse charm dijet cross section is significantly larger than the longitudinal component.  A novel feature is the sharp drop of the longitudinal cross section around $|\mathbf{P}|\sim 1.5$ GeV. A similar, but less drastic, dip is visible for the transverse cross section. 

These minima reflect sensitivity to the size of the projectile $|\mathbf{r}|$ and are understood from Fourier transform to momentum space. A simple estimate of the characteristic inverse 'size' of the photon from its wave function in coordinate space yields $|\mathbf{r}_\gamma|^{-1}\sim \sqrt{m_c^2+z\bar{z}Q^2}\approx 1.4$~GeV, which roughly coincides with the minima of \Fig{fig:IPSat_Pdep_charm}\footnote{The interpretation of these features is  thereby analogous to the origin of minima usually observed in the $t=-|\mathbf{\Delta}|^2$-spectrum of diffractive cross sections. Here, diffractive minima indicate the size of the target, not the projectile. We also note that the light-quark photon wave function is 'larger' than that for charm quarks where we expect a similar feature below 1 GeV (not shown in \Fig{fig:IPSatMagnitude}). We emphasize that the photon is 'non-perturbatively large' for light quarks and small virtuality.}.

\begin{figure}[tb]
\includegraphics[width=0.49\textwidth]{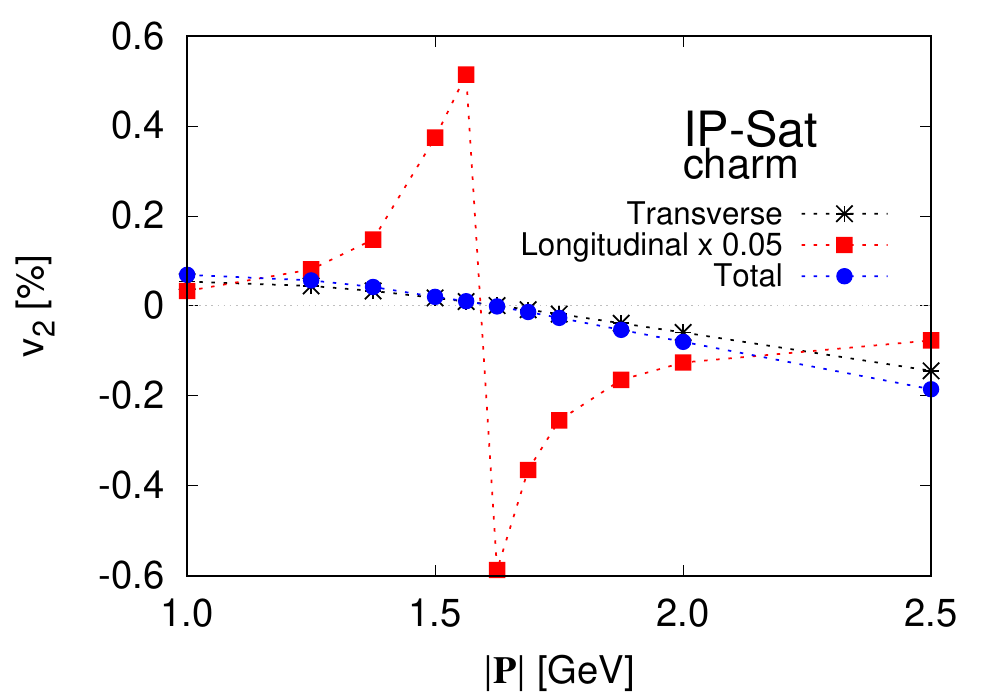}
\caption{$|\mathbf{P}|$-dependence of the elliptic Fourier coefficient $v_2$ for charm jets (integrated over $z\in [0.1,0.9]$), where $|\mathbf{\Delta}| =0.1$ GeV, $Q^2=1$ $\text{GeV}^2$.}\label{fig:IPSatv2_P_charm}
\end{figure}

In \Fig{fig:IPSatv2_P_charm}, we show the $|\mathbf{P}|$ dependence of transverse and longitudinal $v_2$ for charm jets from the modified IP-Sat model with $\tilde{c}=1$. We keep the kinematics as in the light quark case,  $|\mathbf{\Delta}|=0.1$ GeV, $Q^2 =$ 1 GeV${}^2$,  $z\in [0.1,0.9]$. A consequence of the minima shown in \Fig{fig:IPSat_Pdep_charm} is a strong increase of the longitudinal $v_{2,L}$ at $|\mathbf{P}|\sim 1.5$ GeV where 
it changes sign, with $v_{2,L}>0$ for $|\mathbf{P}|\lesssim 1.5$ GeV and $v_{2,L}<0$ for $|\mathbf{P}|\gtrsim 1.5$ GeV. The transverse component, too, changes sign which we attribute to the relative importance of the two terms in \Eq{eq:transvCS}. Specifically, in \Eq{eq:transvCS} the second, mass dependent term dominates for large $|\mathbf{P}|$. Because this term behaves similarly as the corresponding contribution to the longitudinal cross section, it also results in a negative overall $v_2$. 

In the following section, we compute the dipole amplitude directly from the Color Glass Condensate effective theory where angular correlations are included ab-initio
and we will reliably extract the energy dependence of our results.
%
%
%
%%%%%%%%%%%%%%%%%%%%%%%%%%%%%%%%%%%%%%%%%%%%%%%%%%%%%%%%%%%%%%%%%%%%%%%%%%%%%%%%%%%%%%%%%%%%%%%%%%%%%%%%%%%%
%	IP-Glasma
%%%%%%%%%%%%%%%%%%%%%%%%%%%%%%%%%%%%%%%%%%%%%%%%%%%%%%%%%%%%%%%%%%%%%%%%%%%%%%%%%%%%%%%%%%%%%%%%%%%%%%%%%%%%
%
\subsection{CGC computation}\label{sec:fullCGC}
%\subsubsection{Overview}
In this section, we compute dijet cross sections from the CGC, as outlined in Section \ref{sec:fullCGCbasics}. For this quantitative study, we do not show light quark results because of the aforementioned sensitivity to large $|\mathbf{r}|\sim \Lambda_{QCD}^{-1}$  contributions which are not under good control theoretically.
\begin{figure}[tb]
\includegraphics[width=0.49\textwidth]{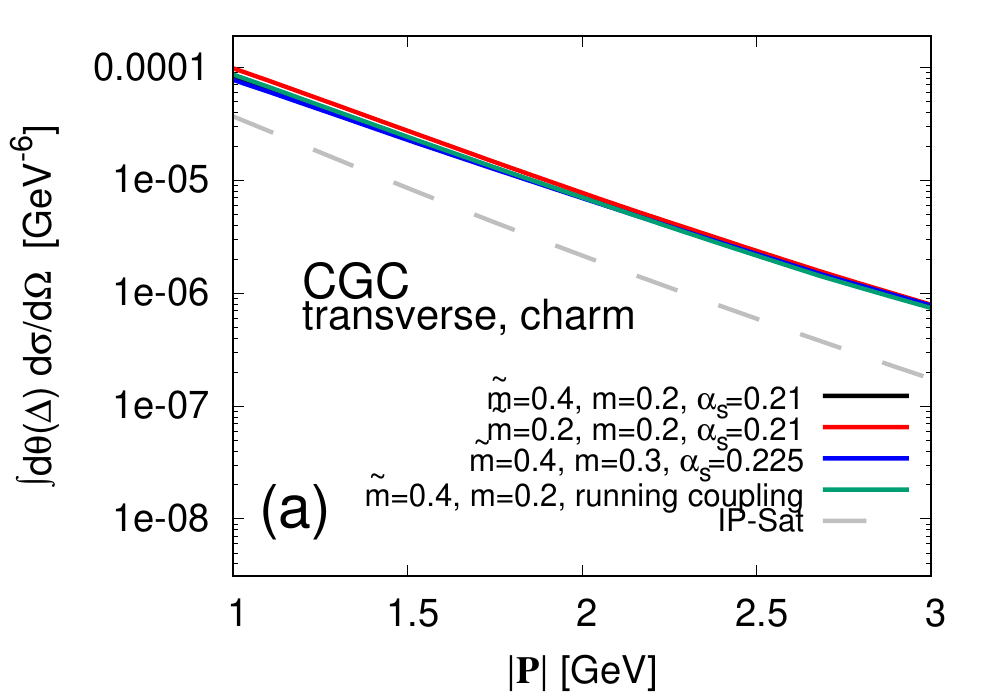}
\includegraphics[width=0.49\textwidth]{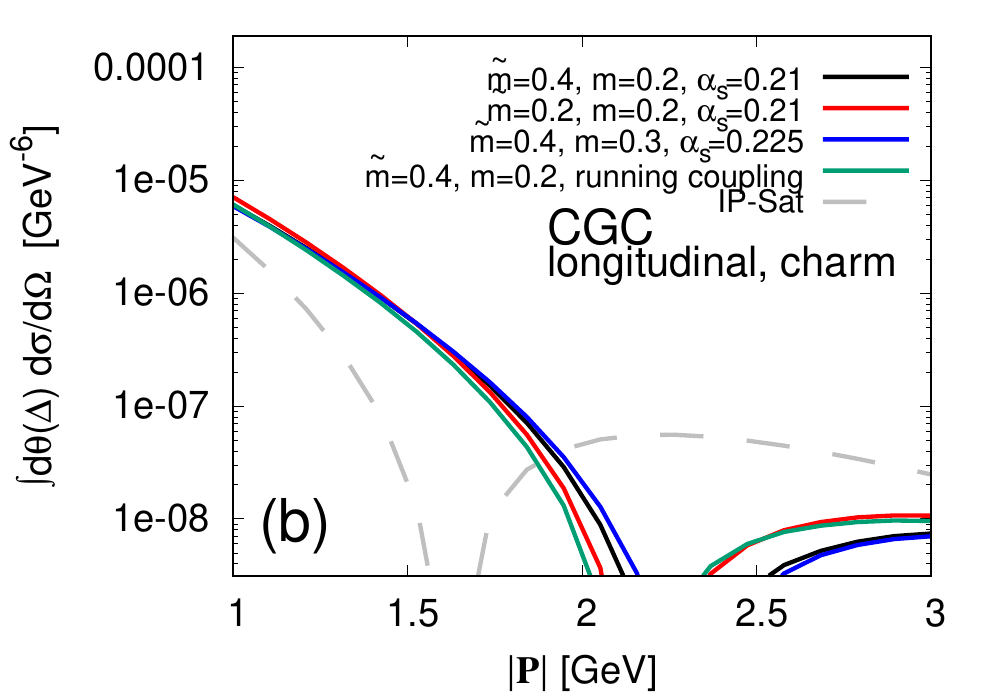}
\includegraphics[width=0.49\textwidth]{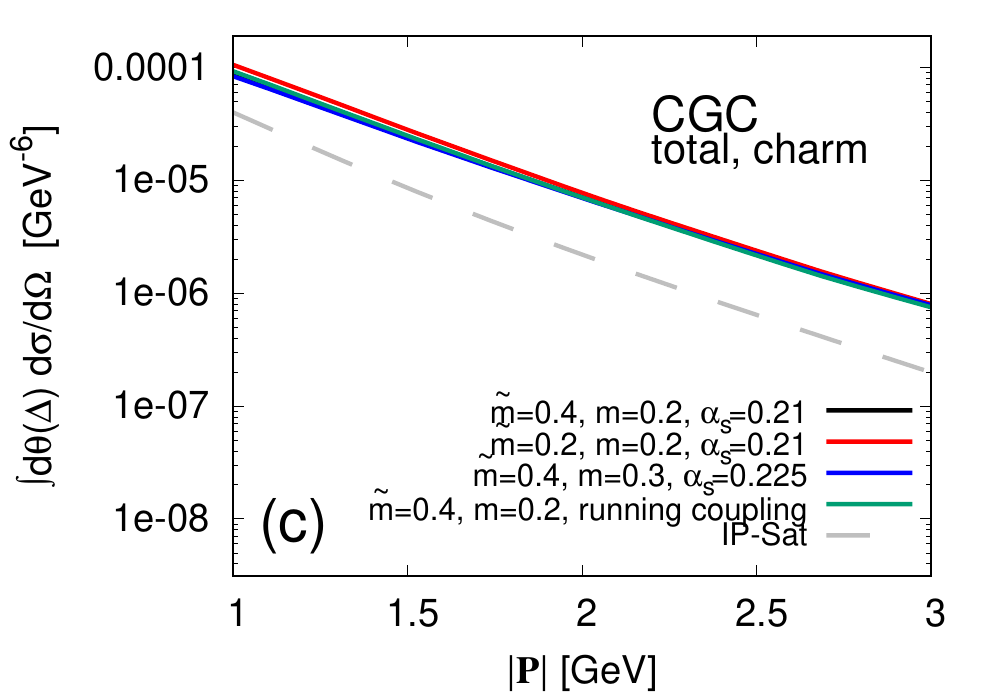}
\caption{Angle integrated transverse (a), longitudinal (b), and total (c) cross sections for charm-dijets from the CGC for $|\mathbf{\Delta}|=0.1$ GeV, $Q^2=1$ GeV${}^2$, $z=\bar{z }=0.5$.}\label{fig:IPGlasma_Pdep_charm_all}
\end{figure}
\begin{figure}[tb]
\includegraphics[width=0.48\textwidth]{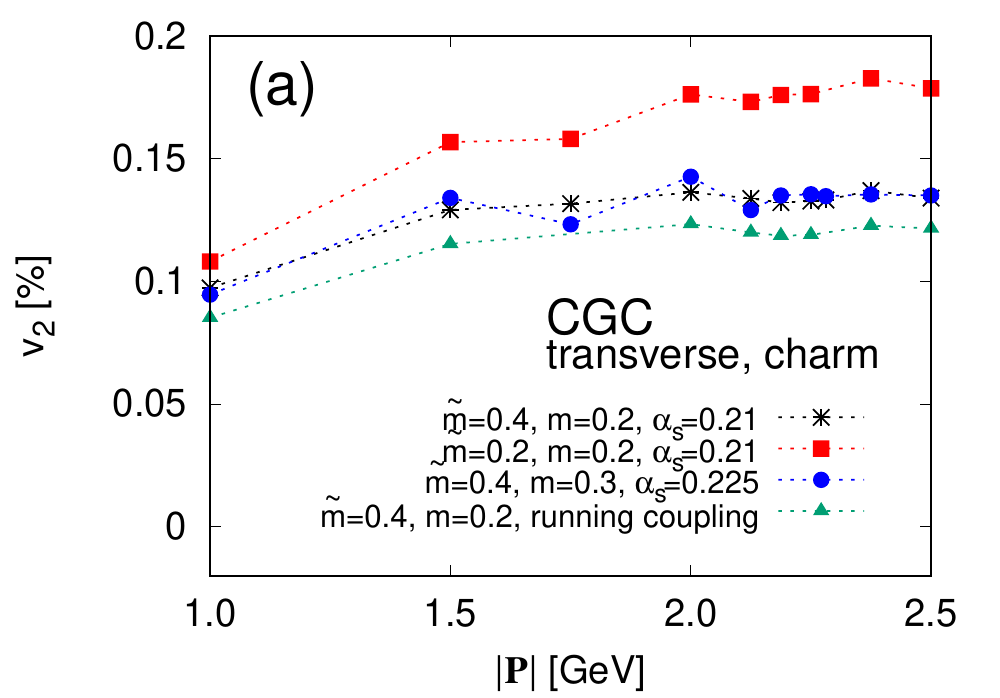}
\includegraphics[width=0.48\textwidth]{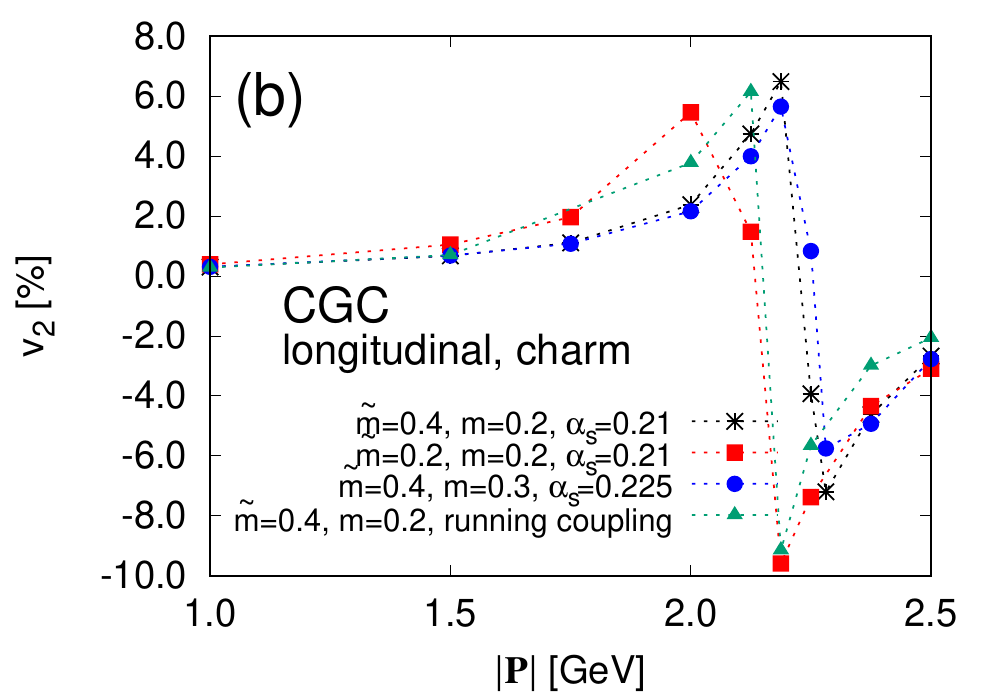}
\includegraphics[width=0.48\textwidth]{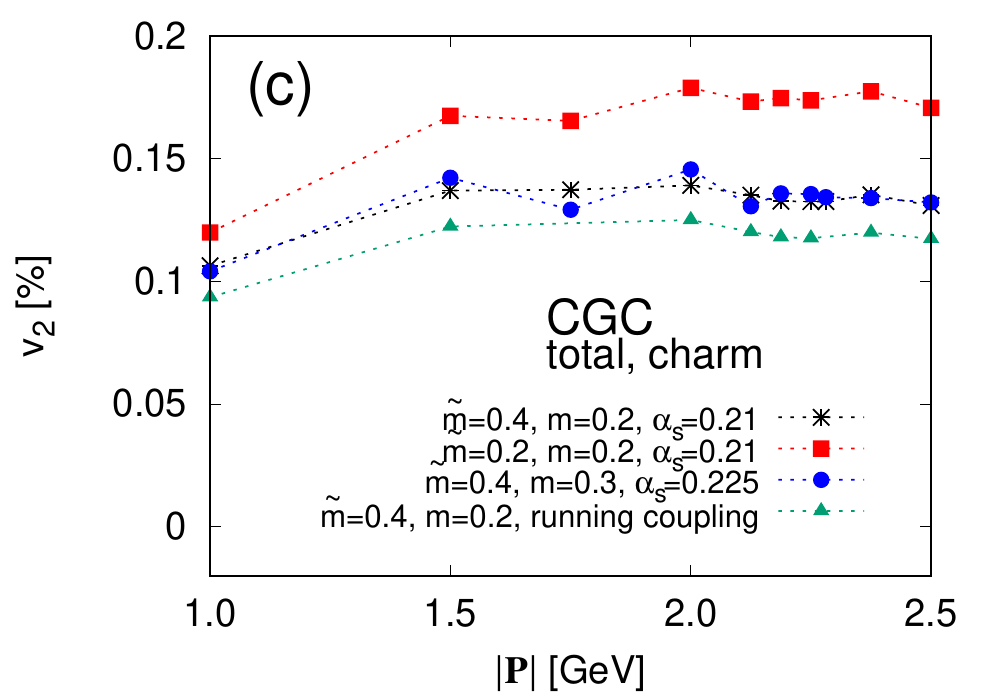}
\caption{Elliptic Fourier coefficients for the charm-dijet cross section for transversely (a) and longitudinally (b) polarized photons from the CGC for $|\mathbf{\Delta}|=0.1$ GeV, $Q^2=1$ GeV${}^2$. The $v_2$ of the total cross section is shown in (c). Results are integrated over $\theta(\mathbf{\Delta})$ and over $z\in [0.1,0.9]$.}\label{fig:IPGlasma_v2_charm_all}
\end{figure}

We plot transverse, longitudinal and total cross sections in Fig.(\ref{fig:IPGlasma_Pdep_charm_all}), where $|\mathbf{\Delta}|=0.1$ GeV, $Q^2 = 1$ GeV${}^2$, $z=\bar{z}=0.5$ and $\theta(\Delta,\mathbf{P})=\pi$.  We vary the infrared regulators $m,\,\tilde{m}$ (see \Eq{eq:YangMillsSolution} and \Eq{eq:JIMWLKkernel}) and the value of a constant coupling $\alpha_s$, as well as a parameterization with running coupling, \Eq{eq:runningcoupling}. We match the value of the saturation scale according to \Eq{eq:impactparam} with $c=0.75$ for an IR-regulator $\tilde{m}=0.4$ GeV, and $c=0.85$ for $\tilde{m}=0.2$ GeV, thus ensuring a consistent overall dipole normalization. We further adjust the value of fixed $\alpha_s$, when we change $m$ in the JIMWLK kernel, \Eq{eq:JIMWLKkernel}, to ensure comparable energy evolution speed. We use this specific parameter range, because it has been used to compute proton structure functions and diffractive vector meson cross sections in \cite{Mantysaari:2018zdd}, and was shown to be consistent with HERA data. 

Shown in Fig.(\ref{fig:IPGlasma_Pdep_charm_all}), the CGC-parameter dependence is very small for charm dijet cross sections, indicating a negligible sensitivity to large distances $|\mathbf{r}|\sim\Lambda_{QCD}^{-1}$.  For comparison we include the IP-Sat results (gray dashed curve), which differ significantly from the CGC in magnitude, but show similar qualitative features. We attribute the different normalization to the fact that in the CGC the small-$|\mathbf{r}|$ regime of the dipole amplitude carries a larger relative weight, compared to IP-Sat, as shown in Fig.(5) of \cite{Mantysaari:2018zdd}. For charm dijets the relevance of this small-$|\mathbf{r}|$ regime is enhanced.

In Fig.~(\ref{fig:IPGlasma_v2_charm_all}) we plot the $v_2$ obtained from the CGC computation for transverse, longitudinal, and total cross sections as a function of $|\mathbf{P}|$, integrated over azimuthal angle $\theta(\mathbf{\Delta})$ and  $z,\bar{z}\in [0.1,0.9]$.  \Fig{fig:IPGlasma_v2_charm_all} (a) shows the transverse $v_{2,T}$ for all CGC parameter sets discussed above. All parameterization agree qualitatively and the results are robust under variation of JIMWLK evolution parameters (stars, circles, and triangles). Including running coupling $\alpha_s$ via \Eq{eq:runningcoupling}, versus  fixed coupling only minimally effects the JIMWLK evolution. Results with different IR-regulator $\tilde{m}$ differ by up to 20 \% (squares versus stars, circles, and triangles). Both values $\tilde{m}=0.2$ GeV and $\tilde{m}=0.4$ GeV are in agreement with HERA data \cite{Mantysaari:2018zdd} and further comparison with data is necessary to reduce this systematic uncertainty.

In \Fig{fig:IPGlasma_v2_charm_all} (b) we show results for the longitudinal anisotropy coefficient $v_{2,L}$, where the most prominent feature is the rapid growth of $v_{2,L}$ in the $|\mathbf{P}|$ region where the longitudinal cross section drops, see \Fig{fig:IPGlasma_Pdep_charm_all} (b). All CGC parameterization give qualitatively and quantitatively similar results, and we attribute the largest uncertainty to the dependence on $\tilde{m}$.

The total $v_2$, which is a weighted sum of transverse and longitudinal components, is shown in  \Fig{fig:IPGlasma_v2_charm_all} (c). At small $|\mathbf{P}|$, longitudinal and transverse cross sections have positive $v_2$ and the total cross section is dominated by the transverse component. At large $|\mathbf{P}|\sim 2.5 $ GeV, the longitudinal $v_{2,L}$ is negative while again the total $v_2$ is dominated by photons with transverse polarization.

\begin{figure}[tb]
\includegraphics[width=0.49\textwidth]{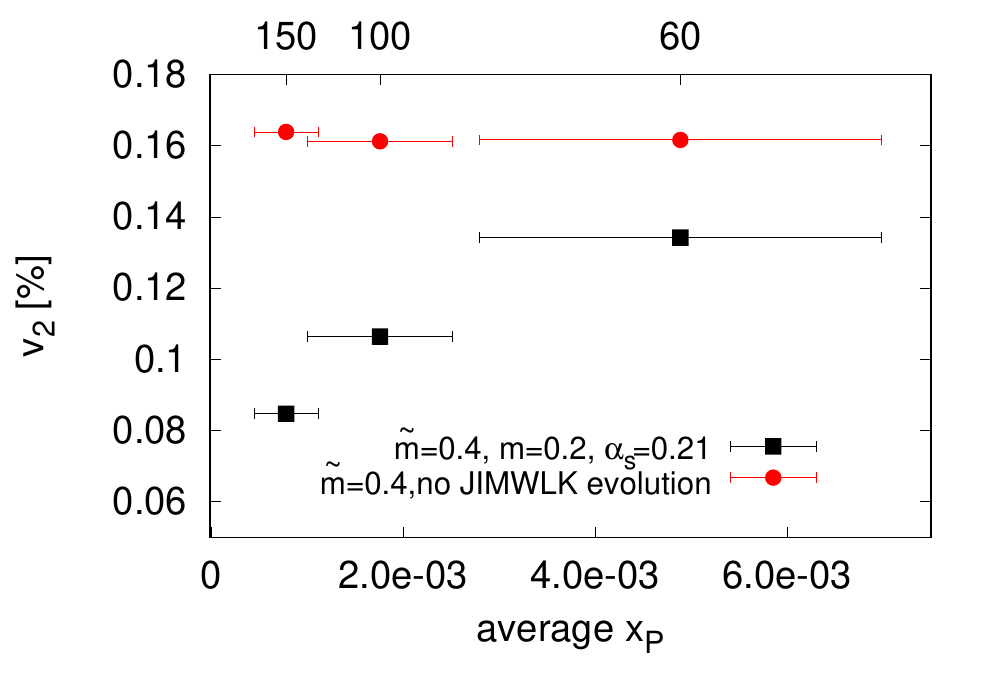}
\caption{Energy dependence of the total $v_2$ for fixed kinematics $|\mathbf{P}|=1$ GeV,  $|\mathbf{\Delta}|=0.1$ GeV, $Q^2=1$ GeV${}^2$ and integrated over $\theta(\mathbf{\Delta})$ and over $z\in [0.1,0.9]$. We compare full CGC results (squares) with results without JIMWLK evolution (circles).}\label{fig:IPGlasma_v2_xdep}
\end{figure}

%
%%% -------------- xP evolution ----------------------------------------------------------------------- 
%
In \Fig{fig:IPGlasma_v2_xdep} we study the energy dependence of $v_2$, by varying the center-of-mass energy $W$, while keeping the dijet kinematics fixed (at $|\mathbf{P}|=1$ GeV, $\mathbf{\Delta}=$ 0.1 GeV, with photon virtuality $Q^2=1$). We integrate over $z\in[0.1,0.9]$ and over the azimuthal angle $\theta(\mathbf{\Delta})$. Shown are results for three different energies $W=60,\,100\,,150$ GeV. Because our results are integrated over $z$ and $\theta(\mathbf{\Delta},\mathbf{P})$, we plot them as a function of the average $x_\mathbb{P}$ in the respective integration range $z\in[0.1,0.9]$ and  $\theta(\mathbf{\Delta},\mathbf{P})\in [0,2\pi]$ (see \Eq{eq:xPomeron}), where
 horizontal bars denote the standard deviation of $x_\mathbb{P}$ in this range. We compare results from the full CGC computation for $\tilde{m}=0.4$, $m=0.2$ with fixed $\alpha_s=0.21$ (squares) to results without JIMWLK evolution (circles), where the $x_\mathbb{P}$ dependence is entirely determined from the IP-Sat parameterization of HERA data. In the latter case, only the saturation scale $Q_s$ changes with energy, while in the full CGC computation also the transverse size of the proton increases. This growth is accompanied with a relative change in the gradients of the target color density. Consequently, we expect a decrease of $v_2$ with energy, as is in fact the case. Remarkably, the same effect is directly observed in the gluon Wigner and Husimi distributions computed in Section \ref{sec:WIgnerfromCGC}. We further examine and validate this interpretation by studying variations of the proton size in Appendix \ref{app:protonsize}.

\begin{figure}[tb]
\includegraphics[width=0.49\textwidth]{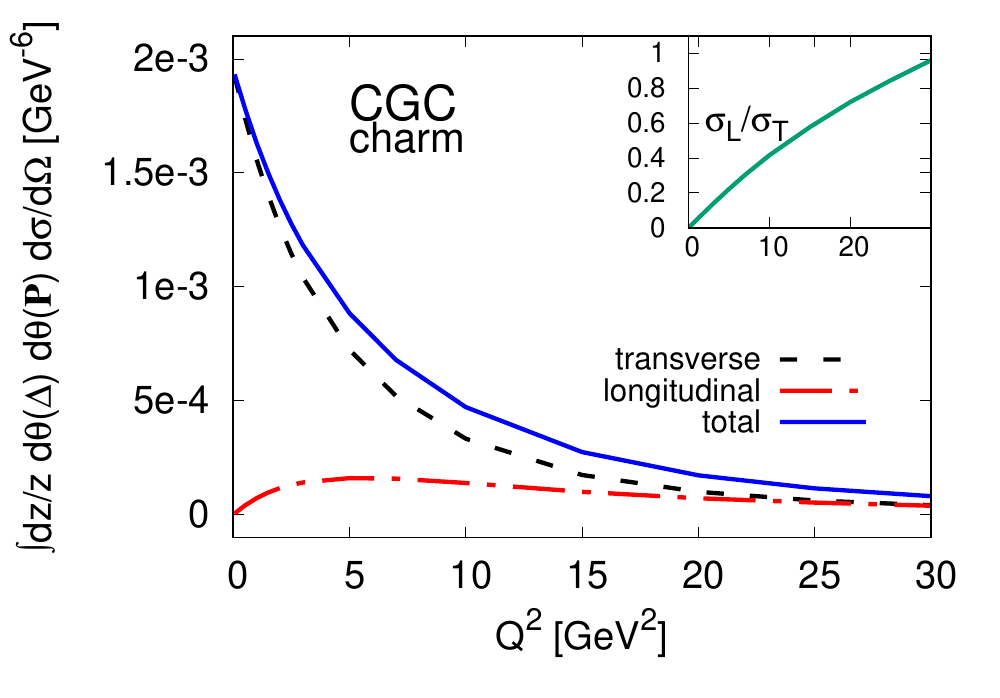}
\caption{Integrated CGC cross sections for charm-dijets as a function of photon virtuality $Q^2$, at $|\mathbf{P}|$=1 GeV, $|\mathbf{\Delta}|$=0.1 GeV, $W=100$ GeV, integrated over the azimuthal angles of the jets, as well as their 
longitudinal momentum $z\in [0.1,0.9]$. Here we use $\tilde{m}=0.4\,{\rm GeV}$, $m=0.2\,{\rm GeV}$ and fixed $\alpha_s=0.21$.}\label{fig:IPGlasma_charm_Q2_CS}
\end{figure}

Finally, we study the dependence of dijet cross section and $v_2$ on photon virtuality $Q^2$.  In \Fig{fig:IPGlasma_charm_Q2_CS} we show  longitudinal and transverse cross sections as a function of $Q^2$, where $|\mathbf{P}|=1$  GeV, $\mathbf{\Delta}=0.1$ GeV and $W=100$ GeV, integrated over $z$  and azimuthal angles, $\theta(\mathbf{\Delta})$ and $\theta(\mathbf{P})$. We choose a CGC parameterization with $m=0.2$ GeV, $\tilde{m}=0.4$ GeV, $c=0.75$ and $\alpha_s=0.21$ (c.f. black curves in \Figs{fig:IPGlasma_Pdep_charm_all}{fig:IPGlasma_v2_charm_all}). While the longitudinal cross section vanishes at $Q^2\sim 0$ (photo-production), it becomes dominant for $Q^2\ge 30  $ GeV${}^2$. Overall, the total cross section decreases at large $Q^2$. The inset shows the ratio of the longitudinal to the transverse cross section.

\begin{figure}[tb]
\includegraphics[width=0.49\textwidth]{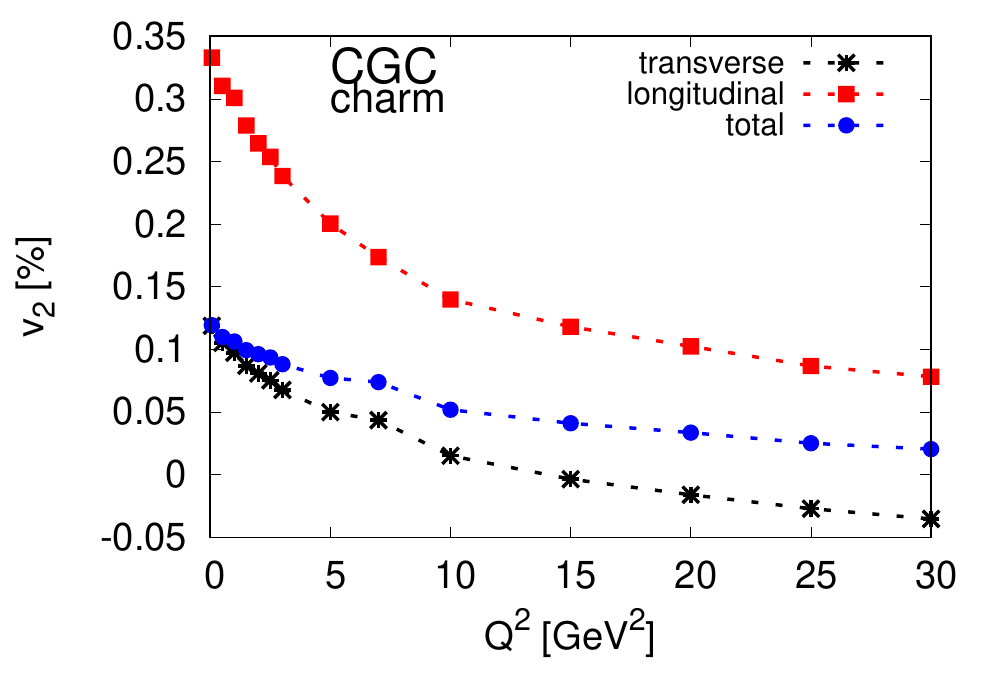}
\caption{Elliptic Fourier coefficient from the CGC computation as a function of photon virtuality.}\label{fig:IPGlasma_charm_Q2_v2}
\end{figure}

In \Fig{fig:IPGlasma_charm_Q2_v2}, we show the $Q^2$-dependence of transverse, longitudinal and total $v_2$ with kinematics as in \Fig{fig:IPGlasma_charm_Q2_CS}. Individually, both transverse and longitudinal $v_2$ decrease with $Q^2$. This is understood, since small dipoles contribute at large $Q^2$ which are less sensitive to correlations in the gluon dipole distribution. Interestingly, the $v_2$ of the transverse component turns negative at approximately $Q^2 = 15\,{\rm GeV}^2$. 
\section{Conclusions}
We have computed gluon Wigner and Husimi distributions of the proton from the Color-Glass-Condensate effective theory, at leading logarithmic order and 
including energy evolution by means of the JIMWLK renormalization group equations. We have studied angular correlations between impact parameter and transverse momentum in Wigner and Husimi distributions, as well as between impact parameter and dipole orientation in the dipole amplitude at small $x$. Our results qualitatively agree with the model calculations of \cite{Hagiwara:2016kam}, but differ at small $|\mathbf{P}|$
where non-perturbative contributions are most important. We studied the energy dependence of azimuthal correlations in gluon distributions, finding a decrease of these correlations with decreasing $x$, related to the geometric growth of the proton.
We investigated differences between Wigner and Husimi distributions and pointed out difficulties for a physical interpretation of the latter due to its dependence on a smearing parameter. 

A possible experimental consequence of elliptic correlations in the gluon Wigner distribution are corresponding elliptic modulations in the diffractive dijet production cross sections in $e+p$ collisions. We computed these within the Color Glass Condensate framework for typical Electron-Ion-Collider energies in the correlation limit and studied angular correlations between dijet transverse momentum and target-recoil. 

Starting from a simple modification of the IP-Sat model, we first established a baseline to disentangle purely kinematical correlations from those originating in the gluon distribution. We found a sizable $\langle \cos{\theta}(\mathbf{\Delta},\mathbf{P})\rangle$ modulation in addition to the expected elliptic contribution. This kinematic effect is due to the energy dependence of the dipole amplitude and vanishes asymptotically in the correlation limit. 
Further, we presented dijet cross sections for charm jets, avoiding  non-perturbative contributions from  large dipoles.

We computed charm dijet cross sections directly from the CGC by solving the JIMWLK renormalization equations for fundamental Wilson lines with initial 
conditions constrained by HERA data at relatively large $x_\mathbb{P}$.  We studied the dependence of the elliptic correlation on relative transverse ($|\mathbf{P}|$) and longitudinal dijet momentum, separately for both photon polarizations.
We predicted the energy dependence of the elliptic anisotropy parameter and found a decrease towards smaller $x$, related to the growth of the proton with energy. 

We computed the $Q^2$ dependence of the dijet cross section and elliptic modulation and observed  a decrease of
 longitudinal and transverse elliptic modulations with photon virtuality, due to the suppression of larger dipoles in the photon wave function with increasing $Q^2$. The transverse $v_{2,T}$ turns negative at approximately $Q^2= 15 $ GeV${}^2$.

Establishing a direct quantitative equivalence between correlations in dijet cross sections and gluon Wigner distributions is possible in restricted cases, such as for dijet production from real ($Q^2\sim 0$) photons in $p+A$ collisions \cite{Hagiwara:2017fye}. In contrast, at the Electron-Ion-Collider one probes a wide range of photon virtualities, beam energies and dijet kinematics. Our study illustrates that many kinematic effects must be considered when trying to extract information
about gluon Wigner and Husimi distributions in the proton from dijet cross sections.  This might be further complicated in a more refined phenomenological analysis, where fragmentation, detector acceptances and other effects are included;
and by parameter dependence of those distributions themselves, as we illustrated for the Husimi distribution.

Nevertheless, in our work we have predicted features of the elliptic modulation of gluon distribution functions, such as its energy dependence, which are manifest in the diffractive dijet cross section and can be easily verified in experiments. The extension of the presented formalism to nuclear targets is straightforward.
 
Constraining the gluon Wigner distribution at small $x$ is also important for future experimental investigation of the proton spin structure at the EIC~\cite{Accardi:2012qut,Lorce:2011kd,Lorce:2011ni,Lorce:2012jy,Leader:2013jra}.
Work to extend the CGC framework to consistently include parton spin in a generalized Wigner function formulation is ongoing \cite{Mueller:2019gjj,Tarasov-Venugopalan-inpreparation}.

\section*{Acknowledgments}
We thank Guillaume Beuf, Renaud Boussarie, Yoshitaka Hatta, Yacine Mehtar-Tani, Alba Soto-Ontoso, Thomas Ullrich, Farid Salazar Wong, Andrey Tarasov and Raju Venugopalan for discussions. 
NM thanks the Department of Physics, University of Jyv\"askyl\"a for their kind hospitality during the completion of this work. HM wishes to thank the Nuclear Theory Group at BNL for hospitality during the early stages of this work. HM is supported by the Academy of Finland, project 314764, and by the European Research Council, Grant ERC-2015-CoG-681707.
NM and BS are supported by the U.S. Department of Energy, Office of Science, Office of Nuclear Physics, under contract No. DE- SC0012704.
NM is funded by the Deutsche Forschungsgemeinschaft (DFG, German Research Foundation) - Project 404640738. This research used resources of the National Energy Research Scientific Computing Center (NERSC), a U.S. Department of Energy Office of Science User Facility operated under Contract No. DE-AC02-05CH11231. Additional computing resources from CSC -- IT Center for Science in Espoo, Finland and from the Finnish Grid and  Cloud Infrastructure (persistent identifier  urn:nbn:fi:research-infras-2016072533) were also used.
\appendix
\section{Wigner functions in the CGC}\label{app:detailsWignerCGC}
In this appendix, we provide details on the computation of gluon Wigner and Husimi distributions from the CGC. The GTMD gluon distribution is defined by \Eq{eq:GTMDgeneral} \cite{Meissner:2009ww,Lorce:2013pza,Echevarria:2016mrc}, 
\begin{align}
x&G_\text{DP}(x,\mathbf{k},\boldsymbol{\Delta}) = 2\int \frac{\der z^- \der^2\mathbf{z}}{(2\pi)^3 \,P^+}e^{-i\mathbf{k}\cdot\mathbf{z}-ixP^+z^-}\nonumber\\
&\times \langle P+\frac{\boldsymbol{\Delta}}{2}| \text{Tr}\left[ F^{+i}\left(\frac{z}{2}\right)\,\mathcal{U}^{-\dagger} F^{+i}\left(-\frac{z}{2}\right)\,\mathcal{U}^{+}  \right] | P-\frac{\boldsymbol{\Delta}}{2}\rangle\,,
\end{align}
where $z=(z^-,\mathbf{z})$. Here, 
\begin{align}
\mathcal{U}^{-\dagger}&\equiv  {U}_{\frac{z^-}{2},-\infty;\frac{\mathbf{z}}{2}}{U}_{-\infty,-\frac{z^-}{2};-\frac{\mathbf{z}}{2} } \,,\\\mathcal{U}^{+}&\equiv  {U}_{-\frac{z^-}{2},\infty;-\frac{\mathbf{z}}{2}} {U}_{\infty,\frac{z^-}{2};\frac{\mathbf{z}}{2} }
\end{align}
are Wilson lines in the fundamental representation. Assuming $xP^+\approx 0$ in the small $x$-regime and using the cyclicity of the trace, we write
\begin{align}\label{eq:GTMD3}
x&G_\text{DP}(x,\mathbf{k},\boldsymbol{\Delta}) \approx 2 \int \frac{\der z^- \der^2\mathbf{z}}{(2\pi)^3 \,P^+}e^{-i\mathbf{k}\cdot\mathbf{z}} \nonumber\\&\qquad\qquad\times \langle P+\frac{\boldsymbol{\Delta}}{2}|\text{Tr}\Big[ \mathcal{U}_{\infty,\frac{z^-}{2};\frac{\mathbf{z}}{2} } F^{+i}\left(\frac{z}{2}\right)  \mathcal{U}_{\frac{z^-}{2},-\infty;\frac{\mathbf{z}}{2}} \nonumber\\&\qquad\qquad\times\mathcal{U}_{-\infty,-\frac{z^-}{2};-\frac{\mathbf{z}}{2} } F^{+i}\left(-\frac{z}{2}\right) \mathcal{U}_{-\frac{z^-}{2},\infty;-\frac{\mathbf{z}}{2}} \Big]| P-\frac{\boldsymbol{\Delta}}{2}\rangle\,.
\end{align}
We now define $\mathbf{z}=\mathbf{x}-\mathbf{y}$ and introduce an irrelevant center coordinate $\mathbf{Z}=(\mathbf{x}+\mathbf{y})/2$, inserting $1=(1/V)\int \der Z^- \der^2\mathbf{Z}$ into \Eq{eq:GTMD3}.
We then use $\der^3Z \der^3z = \der^3x\der^3y$ and obtain
\begin{align}
x&G_\text{DP}(x,\mathbf{k},\boldsymbol{\Delta}) = \frac{4}{2VP^+} \int \der x^- \der y^- \,\der^2\mathbf{x}\,\der^2\mathbf{y} \,e^{-i\mathbf{k}(\mathbf{x}-\mathbf{y})}\nonumber\\&\qquad\qquad\times \langle P+\frac{\boldsymbol{\Delta}}{2}|\text{Tr}\Big[ \mathcal{U}_{\infty,\frac{z^-}{2};\frac{\mathbf{z}}{2} } F^{+i}\left(\frac{z}{2}\right)  \mathcal{U}_{\frac{z^-}{2},-\infty;\frac{\mathbf{z}}{2}} \nonumber\\&\qquad\qquad\times\mathcal{U}_{-\infty,\frac{z^-}{2};-\frac{\mathbf{z}}{2} } F^{+i}\left(-\frac{z}{2}\right) \mathcal{U}_{\frac{z^-}{2},\infty;-\frac{\mathbf{z}}{2}} \Big]| P-\frac{\boldsymbol{\Delta}}{2}\rangle\nonumber\\
\end{align}
Further, using 
\begin{align}
\partial^i U(\mathbf{x})= ig \int\limits_{-\infty}^{\infty}\der z^- U_{-\infty,z^-;\mathbf{x}} \, \partial^i A^+(z^-,\mathbf{x}) U_{z^-,\infty; \mathbf{x}}
\end{align}
and replacing the 
quantum mechanical matrix elements by a stochastic average in the CGC effective theory, $\langle  P+\frac{\boldsymbol{\Delta}}{2}| \dots |  P-\frac{\boldsymbol{\Delta}}{2}| \rangle / (2VP^+) \rightarrow \langle \dots \rangle_x$, yields the final result
\begin{align}
xG_\text{DP}(x,\mathbf{k},\boldsymbol{\Delta}) = \frac{2}{\alpha} \int& \der^2\mathbf{x}\,\der^2\mathbf{y} \,e^{-i\mathbf{k}(\mathbf{x}-\mathbf{y})} \nonumber\\&\times\left( \nabla_{\mathbf{x}}\cdot\nabla_{\mathbf{y}} \right) \langle  \text{Tr} U(\mathbf{x}) U^\dagger(\mathbf{y})\rangle_x\,,
\end{align}
where $\alpha=g^2/4\pi$. A similar derivation is presented in \cite{Dominguez:2011wm}.

An effective method to compute the Wigner distribution numerically was developed in Ref.~\cite{Hagiwara:2016kam}. We include only the first two non-zero harmonic components in the expansion, and write the Wigner distribution as
\begin{equation}
xW(x,\Pt,\bt) = xW_0 + 2 xW_2 \cos(2\theta(\Pt,\bt)).
\end{equation}
The coefficients read
\begin{multline}
xW_0(x,P,b) = -\frac{N_\mathrm{c}}{2\alpha_s \pi^2} \left( \frac{1}{4} \frac{\partial^2}{\partial b^2} + \frac{1}{4b} \frac{\partial}{\partial b} + k^2 \right)  \\
\times \int  \der r r J_0(Pr)  \int_0^{2\pi} \der \theta_{rb} N(r,b, \theta_{rb},x)\,,
\end{multline}
and
\begin{multline}
xW_2(x,P,b) = \frac{N_\mathrm{c}}{2\alpha_s \pi^2} \left( \frac{1}{4} \frac{\partial^2}{\partial b^2} + \frac{1}{4b} \frac{\partial}{\partial b} -\frac{1}{b^2}  + k^2 \right)  \\
\times \int  \der r r J_2(Pr)  \int_0^{2\pi} \der \theta_{rb} \cos(2\theta_{rb}) N(r,b, \theta_{rb},x)\,.
\end{multline}

Similarly, in case of the Husimi distribution defined in Eq.~\eqref{eq:husimi_def}, we use the decomposition
\begin{equation}
xH(\Pt,\bt,x)  = xH_0 +  2 xH_2 \cos(2\theta(\Pt,\bt)).
\end{equation}
Now, following again  Ref.~\cite{Hagiwara:2016kam} the components are
\begin{multline}
xH_0 = -\frac{2N_\mathrm{c}}{l^4 \alpha_s \pi} \int \frac{\der r\,r\, \der b'\, b '}{2\pi}  e^{-\frac{1}{l^2}(b^2+b'^2) - \frac{r^2}{4l^2}} \\
\times \left[ \left\{ \left( \frac{1}{l^4} (b^2+b'^2) + l^2k^2 - \frac{r^2}{4l^2}\right) I_0\left( \frac{2b b'}{l^2} \right) \right. \right.  \\
\left.
\left. - \frac{2bb'}{l^2} I_1\left( \frac{2bb'}{l^2} \right) \right\} J_0(Pr)  - P r I_0\left( \frac{2bb'}{l^2}\right) J_1(Pr) \right] \\
\times \int_0^{2\pi} \der \theta_{rb} N(r,b, \theta_{rb},x).
\end{multline}
and
\begin{multline}
xH_2 = \frac{2N_\mathrm{c}}{l^4 \alpha_s \pi} \int \frac{\der r \,r\, \der b'\, b '}{2\pi}  e^{-\frac{1}{l^2}(b^2+b'^2) - \frac{r^2}{4l^2}} \\
\times \left[ \left\{ \left( \frac{1}{l^4} (b^2+b'^2) + l^2k^2 - \frac{r^2}{4l^2}\right) I_2\left( \frac{2b b'}{l^2} \right) \right. \right.  \\
\left.
\left. - \frac{2bb'}{l^2} I_1\left( \frac{2bb'}{l^2} \right) \right\} J_2(Pr)  - P r I_2\left( \frac{2bb'}{l^2}\right) J_1(Pr) \right] \\
\times \int_0^{2\pi} \der \theta_{rb} \cos(2\theta_{rb}) N(r,b, \theta_{rb},x).
\end{multline}
In our numerical study we use $\alpha_s=0.3$.
\section{Coordinate conventions}\label{app:defkinematics}
\begin{figure}[tb]
\includegraphics[width=0.49\textwidth]{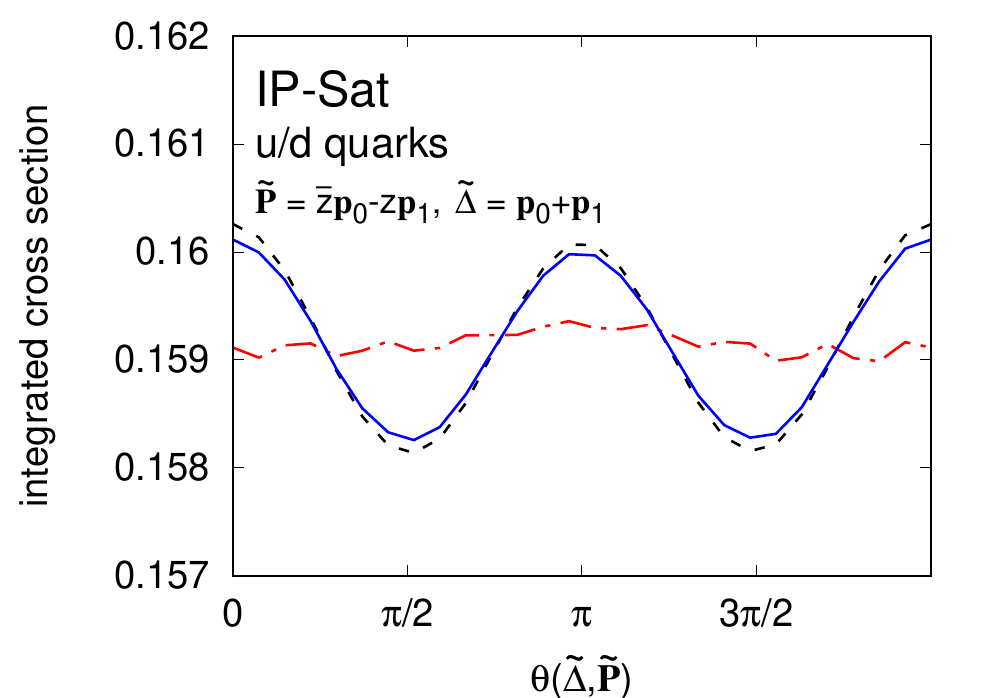}
\caption{Normalized cross sections from the baseline model for $\tilde{c}=0$, \Eq{eq:IPSatwithcorr}, for light quarks as a function of the angle $\theta(\tilde{\mathbf{\Delta}},\tilde{\mathbf{P}})$, with $\tilde{\mathbf{\Delta}}$ $(\tilde{\mathbf{P})}$ defined in \Eq{eq:altdef}. Here, $|\tilde{\mathbf{P}}|$=1 GeV,  $|\tilde{\mathbf{\Delta}}|$=0.1 GeV, $Q^2=$ 1 GeV${}^2$ and integrated over $z\in[0.1,0.9]$. We assume no angular correlations between impact parameter and dipole orientation, yet a non-zero $v_2$ is seen.}\label{fig:alternativedef_overview}
\end{figure}

\begin{figure}[tb]
\includegraphics[width=0.49\textwidth]{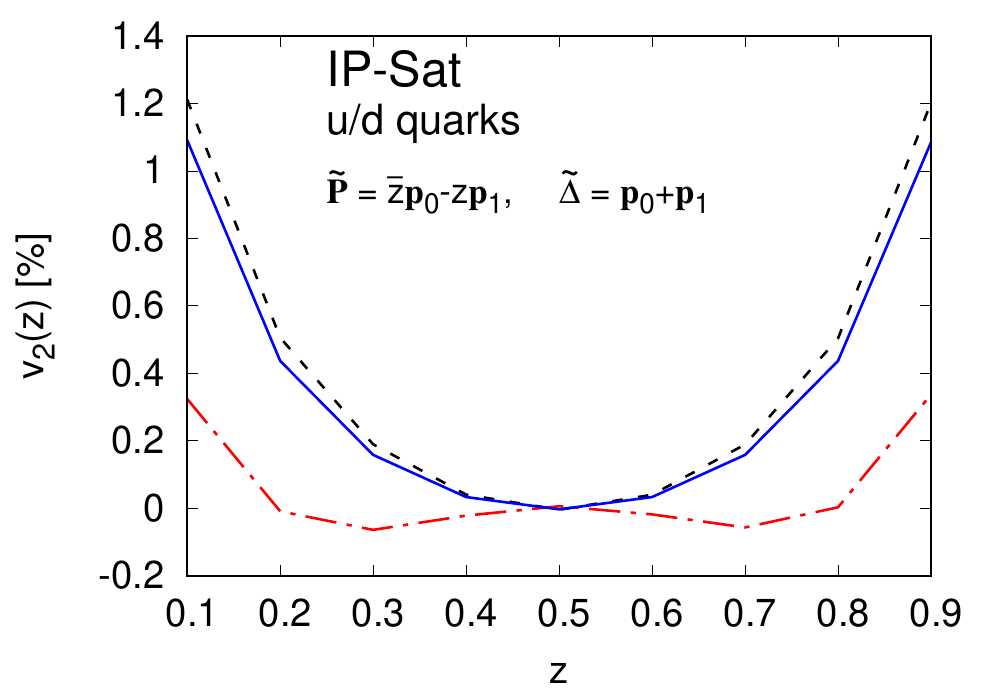}
\caption{$z$-dependence of $v_2$  from the baseline model with $\tilde{c}=0$, \Eq{eq:IPSatwithcorr} with the coordinate definitions $\tilde{\mathbf{\Delta}}$ and $\tilde{\mathbf{P}}$ defined in \Eq{eq:altdef} and parameters as in \Fig{fig:alternativedef_overview}.}\label{fig:alternativedef_v2}
\end{figure}

In this section, we discuss different transverse coordinate definitions to study angular correlations in coherent diffractive dijet production. In the main part of this manuscript we follow the convention of \cite{Hatta:2016dxp} (\Eq{eq:CTRvariables}), but a
different convention is used in \cite{Dumitru:2018kuw} (albeit for a different process),
\begin{align}
\tilde{\mathbf{P}}&\equiv \bar{z}\mathbf{p}_0 - z\mathbf{p}_1\,,\nonumber\\
\tilde{\mathbf{\Delta}}&\equiv \mathbf{p}_0 + \mathbf{p}_1\,.\label{eq:altdef}
\end{align}
For our study, in view of the kinematic effects discussed in Section \ref{sec:baselineIPSat}, an advantage of this convention is that the energy dependence of the dipole amplitude is independent of $\theta(\tilde{\boldsymbol{\Delta}},\tilde{\Pt})$ in these coordinates,
\begin{align}
\tilde{x}_\mathbb{P}=\frac{\frac{1}{z\bar{z}}[m^2+\tilde{\bar{P}}^2] +\mathbf{\Delta}^2+Q^2}{W^2+Q^2-m_N^2}\,.
\end{align}
One may conclude that these variables are more suitable for our study, potentially allowing to extract the gluon distribution at a fixed $x_\mathbb{P}$.
However, now the cross section for dijet production cannot be written as a Fourier integral over the product of photon wave function and dipole distribution. This is because
$\mathbf{r}$, $\mathbf{b}$ and $\tilde{\mathbf{P}}$, $\tilde{\mathbf{\Delta}}$ are not Fourier conjugates. Instead, if we naively evaluate the cross sections \Eq{eq:transvCS} and \Eq{eq:longtCS}
for the IP-Sat model  at fixed $|\tilde{\mathbf{P}}|$, $|\tilde{\mathbf{\Delta}}|$ we find a non-zero $v_2$ even if there are no such correlations on the level of the gluon dipole distribution.
This result is illustrated in \Fig{fig:alternativedef_overview}, where we plot the normalized dijet cross section at  $|\tilde{\mathbf{\Delta}}|=0.1$ GeV, $|\tilde{\mathbf{P}}|=1$ GeV, integrated over $z\in [0.1,0.9]$ and with
photon and proton beam kinematics as in \Fig{fig:IPSatoverviewnormalized}. \Fig{fig:alternativedef_v2} shows the $z$-dependence of the $v_2$ using this convention, where non-zero dijet correlations
are seen even in the absence of correlations in the gluon dipole distribution.

To remedy this situation, one might alternatively parameterize the gluon Wigner distribution in Fourier conjugates to \Eq{eq:altdef},
\begin{align}
\tilde{\mathbf{b}}&=\bar{z}\mathbf{x}_0+z\mathbf{x}_1\nonumber\\
\tilde{\mathbf{r}}&=\mathbf{x}_0-\mathbf{x}_1\,,
\end{align}
thereby mixing longitudinal and transverse coordinates.
%
%	PROTON SIZE
%
\section{Proton size dependence}\label{app:protonsize}
An important consequence of JIMWLK evolution is the growth of the proton with energy. In this appendix we investigate the dependence of $v_2$ on proton size. Here, instead of evolving using the JIMWLK
equations, we directly  match the $x_\mathbb{P}$ dependence of the saturation scale $Q_s$ from the IP-Sat model, c.f. \Eqs{eq:MVgaussian}{eq:impactparam}. In this matching procedure, we manually change the
proton profile \Eq{eq:IPSATprofile} by hand thus mimicking a geometric growth of the target, while keeping the saturation scale fixed.

\begin{figure}[tb]
\includegraphics[width=0.49\textwidth]{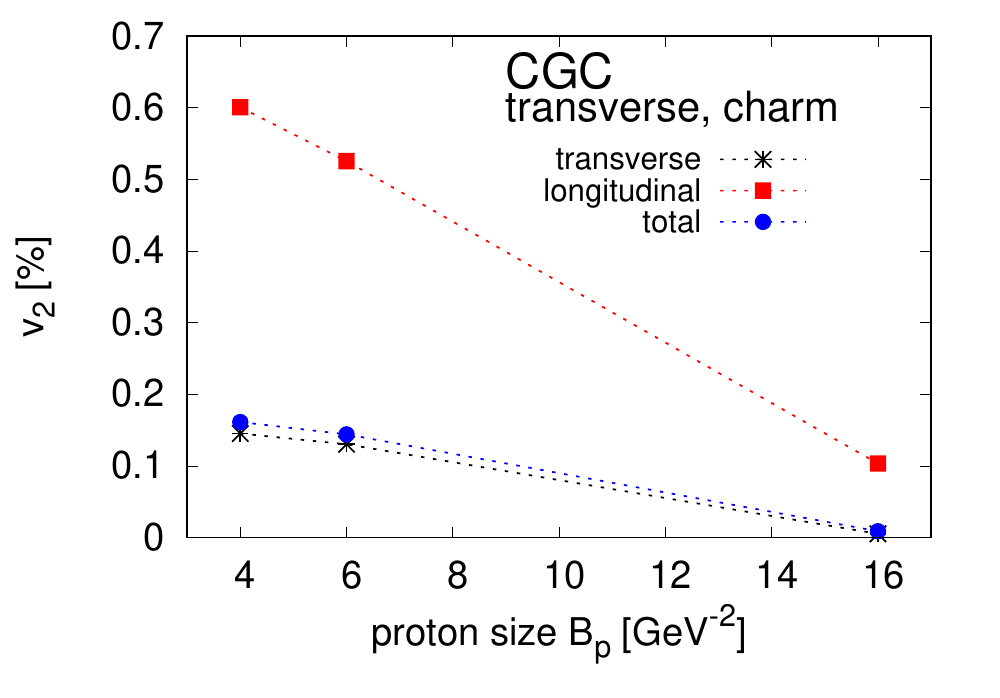}
\caption{Dependence of $v_2$ on the proton size $B_p$ when we match the impact parameter dependence of the CGC computation to the IP-Sat parameterization and change the Gaussian width of the proton by hand.}\label{fig:IPGlasma_charm_protonsize_v2}
\end{figure}

Our results  for $|\mathbf{P}|$=1 GeV, $|\mathbf{\Delta}|$=0.1 GeV, $W=100$ GeV, $Q^2=1$ GeV${}^2$ are shown in \Fig{fig:IPGlasma_charm_protonsize_v2}.  Here we plot the $z$-integrated total $v_2$ as a function
of the proton size in \Eq{eq:IPSATprofile} for $B_p=4,6,16$ GeV${}^{-2}$. For larger $B_p$ the gradients of the color charge density become smoother, thus reducing the dependence of the cross section onto the relative orientation 
between dipole orientation and impact parameter.

These results suggest that the growth of the proton with energy is the most important effect behind the dependence shown in \Fig{fig:IPGlasma_v2_xdep}, while the change of the saturation scale $Q_s$ with energy has a minor effect on $v_2$.

\bibliographystyle{apsrev4-1} 
\bibliography{references}

\end{document}